\documentclass[10pt]{amsart}
\usepackage{amsmath,amssymb,amsthm}
\usepackage{amsmath,amssymb,amsthm,amscd}

\numberwithin{equation}{section}

\title{The Existence of Supersymmetric String Theory with Torsion}

\author{Jun Li}
\author{Shing-Tung Yau}
\thanks{The first author is supported partially by NSF grants
DMS-0200477 and DMS-0244550. The second author is supported
partially by NSF grants DMS-0306600 and DMS-0074329}

\address{Department of Mathematics\\Stanford University\\
Stanford, CA 94305}\email{jli@math.stanford.edu}
\address{Department of Mathematics\\Harvard University\\
Cambridge, MA 02138}\email{yau@math.harvard.edu}
\date{}
\DeclareMathOperator{\spec}{Spec} 
 
\DeclareMathOperator{\End}{End} \DeclareMathOperator{\Ext}{Ext}

\DeclareMathOperator{\Pic}{Pic} 
\DeclareMathOperator{\Endo}{\su\!}

\DeclareMathOperator{\tr}{tr}

\def\L{{\mathbf L}}

\def\T{{\mathbf T}}

\def\cE{{\mathcal E}}
\def\cF{{\mathcal F}}

\def\cH{{\mathcal H}}

\def\cK{{\mathcal K}}
\def\cL{{\mathcal L}}

\def\cO{{\mathcal O}}

\def\cS{{\mathcal S}}

\def\cU{{\mathcal U}}
\def\cV{{\mathcal V}}
\def\cW{{\mathcal W}}

\def\bc{{\mathbf c}}
\def\bB{{\mathbf B}}
\def\be{{\mathbf e}}

\def\bP{{\mathbf P}}
\def\bv{{\mathbf v}}

\def\bA{{\mathbf A}}

\def\bx{{\mathbf x}}

\def\bL{{\mathbf L}}

\def\brr{{\mathbf r}}

\def\bu{{\mathbf u}}
\def\epball{B_\epsilon(T_0)}

\def\ZZ{{\mathbb Z}}

\def\RR{{\mathbb R}}
\def\CC{{\mathbb C}}

\def\HH{{\mathbb H}}

\def\FF{{\mathbf F}}

\def\LL{{\mathbf L}}

\def\uu{{\mathfrak u}}

\def\ii{{I}}

\def\hh{H}

\newtheorem{prop}{Proposition}[section]
\newtheorem{theo}[prop]{Theorem}
\newtheorem{lemm}[prop]{Lemma}

\def\begeq{\begin{equation}}
\def\endeq{\end{equation}}

\def\and{\quad{\rm and}\quad}

\def\Ao{{{\mathbf A}^{\! 1}}}

\def\bl{\bigl(}
\def\br{\bigr)}
\def\lbe{_{\beta}}

\def\cof{\cO_X(5)}
\def\coo{\cO_X(1)}

\def\dual{^{\vee}}

\def\dbar{\bar\partial}
\def\Dbar{D\dpri}

\def\Dbarsta{D^{\prime\prime\ast}}

\def\dpri{^{\prime\prime}}
\def\dds{\frac{d}{ds}}

\def\d{D\pri}
\def\ddt{\frac{d}{dt}}

\def\half{\frac{1}{2}}

\def\her{\mathfrak{Her}^0}

\def\lst{_{S,2}}

\let\lra=\longrightarrow

\def\lbe{_{\beta}}

\def\lalp{_{\alpha}}

\def\lot{_{12}}
\def\lto{_{21}}

\def\lst{_{s,t}}
\def\lzt{_{0,t}}

\def\lij{_{ij}}
\def\lpk{{L^p_k}}
\def\lpkt{{L^p_{k-2}}}
\def\lpko{{L^p_{k-1}}}
\def\lzT{_{0,T}}
\def\lsT{_{s,T}}

\def\mapright#1{\,\smash{\mathop{\lra}\limits^{#1}}\,}

\def\mh{\!:\!}

\def\omz{{\omega_0}}

\def\Omegaoo{\Omega_\RR^{1,1}(X)}

\def\Omegath{\Omega_\RR^{6}(X)}

\def\pri{^{\prime}}

\def\Pfour{{{\mathbf P}^4}}

\def\sub{\subset}
\def\sta{^{\ast}}

\def\sqrtt{\frac{\sqrt{-1}}{2}}
\def\sqrto{{\sqrt{-1}}\,}
\def\su{{\mathfrak su}}

\def\tmetric{\Gamma(\End^0_{\text h}\! E)}

\def\uf{^{\oplus 5}}

\def\upmo{^{-1}}

\def\ucirc{^{\circ}}

\let\uhalf=\uo
\def\uhalf{^{1/2}}
\def\umhalf{^{-1/2}}
\def\ur{^{\oplus r}}
\def\ut{^{\oplus 2}}

\def\lab{\label}

\begin{document}

\maketitle

\section{The system proposed by Strominger}

In their proposed compactification of superstrings \cite{CHSW},
Candelas, Horowitz, Strominger and Witten took the matric product
of a maximal symmetric four dimensional spacetime $M$ with a six
dimensional Calabi-Yau vacua $X$ as the ten dimensional spacetime;
they identified the Yang-Mills connection with the SU(3)
connection of the Calabi-Yau metric and set the dilaton to be a
constant. To make this theory compatible with the standard grand
unified field theory, Witten \cite{W1} and Horava-Witten \cite{HW}
proposed to use higher rank bundles for strong coupled heterotic
string theory so that the gauge groups can be SU(4) or SU(5).
Mathematically, this approach relies on Uhlenbeck-Yau's theorem on
constructing Hermitian-Yang-Mills connections on stable bundles
\cite{YU}. Many authors, including Friedman, Morgan and Witten
\cite{FMW}; Donagi, Ovrat, Pantev and Reinbacher \cite{D4};
Andreas \cite{An}, Kachru \cite{K} and others, have worked on this
subject since then.

In \cite{Str}, A.~Strominger analyzed heterotic superstring
background with spacetime sypersymmetry and non-zero torsion by
allowing a scalar ``warp factor'' to multiply the spacetime
metric. He considered a ten dimensional spacetime that is the
product $M\times X$ of a maximal symmetric four dimensional
spacetime $M$ and an internal space $X$; the metric on $M\times X$
takes the form
$$e^{2D(y)}\left(%
\begin{array}{cc}
g_{ij}(y)&0\\
0&g_{\mu\nu}(x)\\
\end{array}
\right),\qquad x\in X,\quad y\in M;
$$
the connection on an auxiliary bundle is Hermitian-Yang-Mills over
$X$:
$$F\wedge\omega^2=0,\quad  F^{2,0}=F^{0,2}=0
$$
associated to the hermitian form
$\omega=\frac{\sqrt{-1}}{2}g_{i\bar j}dz^i d\bar z^j$. In this
system, following the convention that
$d_c=\sqrt{-1}(\dbar-\partial)$, the physical relevant quantities
are
$$h=\frac{1}{2}d_c\omega,
$$
$$\phi=\frac{1}{8}\log\|\Omega\|+\phi_0,
$$
for a constant $\phi_0$ and
$$g_{ij}^0=e^{2\phi_0}\|\Omega\|^{\frac{1}{4}}g_{ij}.
$$
The spacetime supersymmetry forces $D(y)$ to be the dilaton field.

In order for such ansatze to provide a supersymmetric
configuration, one introduces a Majorana-Weyl spinor $\epsilon$ so
that
$$
\delta \phi_j^0= \nabla^0_j\epsilon^0 +\frac{1}{48}e^{2\phi}\bl
\gamma_j^0 H^0-12 h_j^0\br \epsilon^0=0,
$$
$$
\delta\lambda^0 = \nabla^0\phi
\epsilon^0+\frac{1}{24}e^{2\phi}h^0\epsilon^0=0,
$$
and
$$
\delta \chi^0 = e^{\phi} F_{ij}\Gamma^{0ij}\epsilon^0=0.
$$
Here $\psi^0$ is the gravitano, $\lambda^0$ is the dilatino,
$\chi^0$ is the gluino, $\phi$ is the dilaton and $h$ is the
Kalb-Ramond filed strength obeying
$$dh=\tr R\wedge R-\tr F\wedge F.
$$
(For details of this discussion, please consult \cite{Str, Str2}.)
By suitably transforming these quantities, Strominger showed that
in order to achieve space-time supersymmetry the internal six
manifold $X$ must be a complex manifold with a non-vanishing
holomorphic three form $\Omega$; the Hermitian form $\omega$ must
obey
$$\partial\dbar\omega=\sqrt{-1}\tr F\wedge
F-\sqrt{-1}\tr R\wedge R
$$
and
$$d\sta\omega=d_c\log\|\Omega\|.
$$
Accordingly, he proposed to solve the system
\begin{equation}\lab{0.1}
F\wedge\omega^2=0;
\end{equation}
\begin{equation}\lab{0.2}
F^{2,0}=F^{0,2}=0;
\end{equation}
\begin{equation}\lab{0.3}
\partial\dbar \omega=
\sqrt{-1}\tr F\wedge F-\sqrt{-1}\tr R\wedge R;
\end{equation}
and
\begin{equation}\lab{0.4}
d\sta \omega=d_c\log\|\Omega\|.
\end{equation}
These are solutions of superstrings with torsions that allows
non-trivial dilaton fields and Yang-Mills fields\footnote{The
equation (\ref{0.3}) in \cite{Str} has $\frac{1}{30}\tr F\wedge
F$; this is because he worked with principle bundles and the trace
is that of its adjoint bundle.}. Here $\Omega$ is a nowhere
vanishing holomorphic three form on the complex threefold $X$;
$\omega$ is the Hermitian form and $R$ is the curvature tensor of
the Hermitian metric $\omega$; $F$ is the curvature of a vector
bundle $E$; and $\tr$ is the trace of the endomorphism bundle of
either $E$ or $TX$.

In \cite{Str}, Strominger found some solutions to this system for
U(1) principle bundles. In this paper we shall give the first
irreducible non-singular solution of the supersymmetric system of
Strominger for U(4) and U(5) principle bundles. We obtain our
solutions by perturbing around the Calabi-Yau vacua paired with
the gauge field on the tangent bundle of $X$.

In more concrete term, we take a smooth Calabi-Yau threefold
$(X,\omega)$ and a reducible Yang-Mills connection (metric) $H$ on
$TX\oplus \CC_X\ur$; $(\omega, H)$ is a reducible solution to
Strominger's system. For any small deformations $\dbar_s$ of the
holomorphic structure of $TX\oplus\CC_X\ur$, we derive a
sufficient condition for (\ref{0.1})-(\ref{0.4}) being solvable
for $(X,\dbar_s)$: it is that the Kodaira-Spencer class of the
family $\dbar_s$ at $s=0$ satisfies certain non-degeneracy
condition (see Theorem \ref{p4.2}). After that, we will construct
examples of Calabi-Yau threefolds that admit small deformations of
$TX\oplus\CC_X\ur$ satisfying this requirement. This provides the
first example of regular irreducible solution to Strominger's
system.

In the next paper, we would like to understand the
non-perturbative theory and hope to formulate a global structure
theorem of the moduli space of these fields.

It was speculated by M.~Reid that all Calabi-Yau manifolds can be
deformed to each other through conifold transition. To achieve
this goal, it is inevitable that we must work with non-Kahler
manifolds. We hope that such non-Kahler manifolds will adopt the
Strominger structures. We shall come back to this in the second
paper.

%ch1.tex

\section{Solving Hermitian-Einstein equation by perturbation}

In this section we will solve the usual Hermitian-Yang-Mills
system using perturbation method. Let $(E,\Dbar_s)$ be a smooth
family of holomorphic vector bundles on an $n$-dimensional Kahler
manifold $(X,\omega)$. Suppose $\hh_0$ is a Hermitian-Yang-Mills
metric on $(E,\Dbar_0)$; we ask under what condition can we extend
$\hh_0$ to a smooth family of Hermitian-Yang-Mills metrics $\hh_s$
on $(E,\Dbar_s)$? When $\hh_0$ is irreducible, the answer is
affirmative. The case when $\hh_0$ is reducible is more subtle.
Let $(E_1,\Dbar_1)$ and $(E_2,\Dbar_2)$ be two degree zero
slope-stable vector bundles on $X$. By a theorem of Uhlenbeck-Yau,
both $(E_1,\Dbar_1)$ and $(E_2,\Dbar_2)$ admit
Hermitian-Yang-Mills metrics $\hh_1$ and $\hh_2$. The direct sum
of their scalar multiples $\hh_1\oplus e^t \hh_2$ is a
Hermitian-Yang-Mills metric on
$$(E,D\dpri_0)\triangleq (E_1\oplus E_2,\Dbar_1\oplus\Dbar_2).
$$
Suppose we are given a smooth deformation of holomorphic
structures $\Dbar_s$ of $\Dbar_0$, then the Kodaira-Spensor class
identifies the first order deformation of the family $\Dbar_s$ at
$0$ to an element
$$\kappa\in H^1_{\dbar}(X,E\dual\otimes E)
$$
in the Dolbeault cohomology of the $\dbar$-operator $\Dbar_0$.
Because $\Dbar_0=\Dbar_1\oplus\Dbar_2$, the above cohomology space
decomposes into direct sum
$$\oplus_{i,j=1}^2 H^1_{\dbar}(X,E_i\dual\otimes E_j).
$$
We let $\kappa_{ij}\in H^1_{\dbar}(X,E_i\dual\otimes E_j)$ be its
associated components under this decomposition.

\begin{theo}\lab{p2.1}
Suppose $\kappa\lot$ and $\kappa\lto$ are non-zero, then there is
a unique $t$ so that for sufficiently small $s$ the metric
$\hh_0(t)=\hh_1\oplus e^t \hh_2$ extends to a family of
Hermitian-Yang-Mills-metrics $\hh_s$ on $(E,\Dbar_s)$.
\end{theo}

We will prove this theorem by applying implicit function theorem
to the elliptic system of the Hermitian-Yang-Mills metrics of
$(E,\Dbar_s)$.

To begin with, equation (\ref{0.2}) holds for any hermitian
connections of holomorphic vector bundles. Now let $H$ be a
hermitian metric on $E$ and $F_{s,H}$ be its the hermitian
curvature on $(E,\Dbar_s)$. The Hermitian-Yang-Mills equation for
$(E,\Dbar_s)$, which has degree zero, then becomes
\begin{equation}\lab{2.25}
F_{s,H}\wedge\omega^{n-1}=0.
\end{equation}
The linearization of (\ref{2.25}) is self-adjoint and has
two-dimensional kernel and cokernel. In case
$\wedge^r(E,\Dbar_0)\cong \CC_X$, we can normalize $H$ so that its
induced metric on $\wedge^r E\cong\CC_X$ is the constant one
metric. Then the linearization of the restricted system has one
dimensional kernel and cokernel. We suppose the cokernel is
spanned by $J\cdot\omega^n$. Then for small $s$, the implicit
function theorem supplies us a one dimensional family of solutions
$\hh\lst$, of determinant one, to the system (\ref{2.25}) modulo
the linear span of $J\cdot\omega^n$:
\begin{equation}\lab{0.21}
F_{s,\hh\lst}\wedge \omega^{n-1}\equiv 0\mod J\cdot\omega^n.
\end{equation}
To prove the theorem, it remains to show that we can find a
function $t=\rho(s)$ so that
\begin{equation}\lab{0.22}
F_{s,\hh_{s,\rho(s)}}\wedge \omega^{n-1}=0.
\end{equation}
For this, we will look at the functional
$$r(s,t)=\frac{\sqrt{-1}}{2}\int\tr\bl F_{s,\hh\lst}\cdot
J\br\wedge\omega^{n-1}
$$
and investigate the derivatives $\dot r(0,t)=
\frac{d}{ds}r(s,t)|_{s=0}$. Since $r(0,t)\equiv 0$, the first
order derivatives $\dot r(0,t)$ are independent of the
parameterizations $(s,t)$. By a direct calculation, they all
vanish. Thus we are forced to work at the second order derivatives
$\ddot r(0,t)$; they are of the form
$$\ddot r(0,t)=e^{-\alpha t}A-e^{\alpha t}B,\quad A, B\geq 0.
$$
In case $\kappa\lot$ and $\kappa\lto$ are non-zero, $A$ and $B$
become positive; hence we can find a function $t=\rho(s)$ so that
$\lim_{s\to 0}\rho(s)=\frac{1}{2\alpha}\ln (A/B)$ and
$$r(s,\rho(s))=0.
$$
This shall prove the existence theorem. Since later we will adopt
this approach to solve Strominger's system, we shall provide its
detail here as a warm up.

We begin with the basic objects: the vector bundle, its
holomorphic structure and its curvature. We let $(X,\omega)$ be a
Kahler manifold of dimension $n$; we let $(E_1,D_1\dpri)$ and
$(E_2,D\dpri_2)$ be two degree zero slope stable holomorphic
vector bundles of ranks $r_1$ and $r_2$; we let $<\,,>_1$ and
$<\,,>_2$ be the Hermitian-Yang-Mills metrics of $(E_1,\Dbar_1)$
and $(E_2,\Dbar_2)$. For simplicity, we assume
$\wedge^{r_i}(E_i,\Dbar_i)\cong \CC_X$ and pick $<\,,>_i$ so that
its induced metric on $\wedge^{r_i}E_i\cong\CC_X$ is the constant
$1$ metric. Under this arrangement the $<\,,>_i$ are unique. We
then let $E=E_1\oplus E_2$, of rank $r=r_1+r_2$, and endow it with
the holomorphic structure $\Dbar_1\oplus\Dbar_2$ and the reference
hermitian metric $<\,,>=<\,,>_1\oplus <\,,>_2$.

Next we let $\Dbar_s$ be a smooth family of holomorphic structures
on $E$ so that $\Dbar_0=\Dbar_1\oplus\Dbar_2$. $\Dbar_s$ relates
to $\Dbar_0$ by a global section $A_s\in \Omega^{0,1}\bl\End
E\br$:
$$\Dbar_s=\Dbar_0+A_s;
$$
the hermitian connection $D_s=D\pri_s+\Dbar_s$ of
$(E,D_s\dpri,<,>)$ relates to the hermitian connection $D_0$ of
$(E,D_0\dpri,<,>)$ via
$$D_s=(\Dbar_0+A_s)+(\d_0-A_s\sta);
$$
the hermitian curvature of $D_s$ becomes
\begin{equation}\lab{2.11}
F_s=F_0+(\Dbar_0+\d_0)(A_s-A_s\sta)-(A_s-A_s\sta)\wedge(A_s-A_s\sta).
\end{equation}
Here $A_s\sta$ is the hermitian adjoint of $A_s$ under $<\,,>$.

It is instructive to express them in local coordinates. Let
$e_1,\cdots,e_n$ be a (local) orthonormal basis of $(E, <\,,>)$.
We define the connection form $\Gamma_{s,\alpha\beta}$ of
$D_s\dpri$:
$$\Dbar_s e\lalp=\Gamma_{s,\alpha\beta}e\lbe;
$$
then the matrix
$$A_s=(A_{s,\alpha\beta})=
(\Gamma_{s,\alpha\beta}-\Gamma_{0,\alpha\beta}).
$$
For any local section $v=\sum x\lalp e\lalp$, written in the
matrix form $v=\bx\,\be^t$ with $\bx=(x_1,\cdots,x_n)$ and
$\be=(e_1,\cdots,e_n)$ being row vectors, the differentiation
$$\Dbar_s v=\bl\dbar\bx+\bx(\Gamma_{s,\alpha\beta})\br\be^t
=(\dbar\bx+\bx(\Gamma_{0,\alpha\beta}))\be^t + (\bx A_s)\be^t
=\Dbar_0 v+ v A_s.
$$
In case $\varphi$ is a section of $E\dual\otimes E$, a local
computation shows that
$$\Dbar_s\varphi=\Dbar_0\varphi-[A_s,\varphi]\and
D\pri_s\varphi=D_0\pri\varphi+[A_s\sta,\varphi].
$$
This works for endomorphism-valued $p$ and $q$-forms $A$ and $B$
if we follow the convention $[A,B]=A\wedge B-(-1)^{pq} B\wedge A$.

\begin{lemm}\lab{1.31}
Let $(E,\Dbar_s)$ be a family of holomorphic structures on a
vector bundle $E$ over a Kahler manifold $(X,\omega)$. Then there
is a family of gauge transformations $g_s\in \Omega^0(\End E)$,
$g_0=\text{id}$, so that the first order derivative $\dds g_s\sta
\Dbar_s$ is $\Dbar_0$-harmonic.
\end{lemm}

\begin{proof}
First, we can find $\mu\in\Omega^0(\End E)$ so that $\dot
D\dpri_0+\Dbar_0\mu$ is $\Dbar_0$-harmonic. We then choose a
family of gauge transformation $g_s$ so that $\dds
g_s\sta\Dbar_s=\dot D\dpri_0+\Dbar_0 \mu$. $g_s$ is the desired
family of gauge transformations.
\end{proof}

As a corollary, we can choose the family $\Dbar_s=\Dbar_0+A_s$ so
that $\dot A_0$ is $\Dbar_0$-harmonic with respect to the Kahler
form $\omega$.

Solving Hermitian Yang-Mills connections involves working with
other hermitian metrics of $E$. We let $\cH(E)_1$ be the space of
all hermitian metrics on $E$ whose induced metrics on $\wedge^r
E\cong\CC_X$ are the constant one metric. Once we have the
reference metric $<\,,>$,  the space $\cH(E)_1$ is isomorphic to
the space of determinant one pointwise positive definite
$<\,,>$-hermitian symmetric endomorphisms of $E$ via
$$<\!u,v\!>_H=<\!uH,v\!>.
$$
In this paper, we shall use such $H$ to represent its associated
hermitian metric.

Given a hermitian metric $H$, its hermitian connection $D_{s,\hh}$
is
$$D_{s,\hh}=(D_s\pri+D_s\pri\hh\cdot \hh\upmo)+D\dpri_s;
$$
its curvature is
$$F_{s,\hh}=F_s+\Dbar_s(D\pri_s \hh\cdot \hh\upmo).
$$
The Hermitian-Yang-Mills connections of $(E,\Dbar_s)$ are
hermitian metrics $H\in\cH(E)_1$ making
$$\tilde L_s(\hh)=\bl F_s+\Dbar_s(D\pri_s \hh
\cdot
 \hh\upmo)\br\wedge\omega^{n-1}
$$
vanish. Because $H$ induces the constant one metric on $\wedge^r
E$, $\tr F_{s,H}$, which is the curvature of $(\wedge^r E,\det
H)$, is zero. Hence $\tilde L_s(\hh)$ is traceless $H$-hermitian
antisymmetric. To make it $<\,,>$-hermitian anti-symmetric
instead, we form the operator
\begin{equation}
\lab{L} L_s(\hh)=H^{-1/2}\cdot\tilde L_s(\hh)\cdot H^{1/2}:
\cH(E)_1\lra \Omega^{2n}_\RR(\Endo E).
\end{equation}
It takes value in the vector bundle $\su E$ of traceless hermitian
anti-symmetric endomorphisms of $(E,<\,,>)$.

Next we let $I_i$ be the identity endomorphism of $E_i$, viewed as
an endomorphism of $E$. Because both $E_1$ and $E_2$ are degree
zero slope stable and $H_1$ and $H_2$ are their
Hermitian-Yang-Mills metrics, the solutions to $L_0(\hh)=0$ are
\begin{equation}\lab{ht}
 \hh_{0,t}=e^{t/r_2}
 I_1\oplus e^{-t/r_2} I_2, \quad t\in\RR.
\end{equation}
Further, using $\delta \hh= \hh_{0,t}\umhalf \delta h
\hh_{0,t}\umhalf$, which is an isomorphism of the tangent space of
$\cH(E)_1$ at $H\lzt$ with the space of sections of the vector
bundle $\her E$ of traceless hermitian symmetric endomorphisms of
$(E,<\,,>)$, the linearization of $L_0$ at $ \hh_{0,t}$ becomes
\begin{equation}
\delta L_0(\hh_{0,t})(\delta h)= \Dbar_0D\pri_0\delta h \wedge
\omega^{n-1}: \Omega^0(\her E)\lra \Omega^{2n}_\RR(\Endo E).
\end{equation}

Because $(E,\Dbar_0)$ is a direct sum of two distinct stable
vector bundles, the kernel and the cokernel of $\delta L_0$ are
both one-dimensional, of which the cokernel is spanned by
$$J=\frac{\sqrto}{r_2}\ii_1\cdot\omega^n\oplus
-\frac{\sqrto}{r_1}\ii_2\cdot\omega^n,
$$
independent of $t$. To apply the implicit function theorem, we
take the projection
$$P: \Omega^{2n}_\RR(\Endo E)\lra \Omega^{2n}_\RR(\Endo E)
/\RR\cdot J
$$
and look at the composite
$$P\circ L_s: \cH(E)_1\lra \Omega^{2n}_\RR(\Endo E)/\RR\cdot J.
$$
Because the cokernel of $P\circ \delta L_0$ at $\hh\lzt$ is $0$,
for small $s$ the operator $P\circ L_s$ is an open operator near
$\hh\lzt$. Further because the linearization of $P\circ L_s$ has
index one at $\hh\lzt$, the solution space $V_s$ of $P\circ L_s=0$
is a one-dimensional smooth manifold near $H\lzt$ and the union
$\cup_s V_s$ is a smooth two dimensional manifold near $H\lzt$.
Since $V_0$ is parameterized by the line $\RR$ via the solutions
(\ref{ht}), we can extend this parameterization to $V_s$ near
$H\lzt$ so that $(s,t)$ provides a coordinate chart of $\cup_s
V_s$. We let $ \hh_{s,t}$ be the solution to $P\circ L_s=0$
associated to $(s,t)\in V_s$. This way, to solve $L_s(\hh)=0$ if
suffices to find the vanishing loci of the function
$$r(s,t)=\sqrto\int_X \tr \bl L_s(\hh_{s,t})\cdot\ii_1\br\in\RR.
$$

We will show that there is a function $t=\rho(s)$ so that
$r(s,\rho(s))=0$. Because $r(0,t)=0$, the first step is to
investigate the sign of the derivatives of $r(s,t)$ of $s$ at
$s=0$. Recall that
$$\Dbar_s \hh\lst=\Dbar_0 \hh\lst-[A_s, \hh\lst]\and
\d_s \hh\lst=\Dbar_0 \hh\lst+[A_s\sta, \hh\lst];
$$
hence %the first and the second order derivatives
$$\frac{d}{ds}\,\Dbar_s \hh\lst=\Dbar_s \dot \hh\lst-[\dot A_s, \hh\lst]
\and \frac{d}{ds}\,\d_s \hh\lst=\d_s \dot \hh\lst+[\dot A_s\sta,
\hh\lst].
$$
Therefore, following the convention that $\dot f(s,t)=\dds f(s,t)$
and $\dot f(0,t)=\dot f(s,t)|_{s=0}$,
$$\dds L_s(\hh\lst)=\dot F_s-[\dot A_s,D _s
\hh\lst\cdot \hh\lst\upmo]+\Dbar_s\varphi\lst
\quad\text{with}\quad\varphi\lst=\dds (\d_s \hh\lst\cdot
\hh\lst\upmo).
$$

We have the following useful easy observation:

\begin{lemm}\lab{2.2}
Let $\mu_1\in \Omega^{1,0}(E_1\dual\otimes E_2)$, let $\mu_2\in
\Omega^{0,1}(E_1\dual\otimes E_2)$, and let
$\psi\in\Omega^0(E_2\dual\otimes E_1)$ be a smooth section.
\newline
1. Suppose $\Dbar_0\psi=0$, then
$\displaystyle\int_X\tr(\Dbar_0\mu\cdot\psi)\wedge\omega^{n-1}=0$.
\newline
2. Suppose $\Dbarsta_0 \mu_2=0$, then $\displaystyle
\int_X\tr(\mu_2\cdot\d_0\psi)\wedge\omega^{n-1}=0$.
\end{lemm}

\begin{proof}
The two identities follow directly from the Stokes' formula.
First, because $\Dbar_0 \psi=0$, because $\mu$ is a $(1,0)$-form
and because $\omega$ is a Kahler form on $X$,
$$\int_X\tr(\Dbar_0\mu_1\cdot\psi)\wedge\omega^{n-1}=
\int_X\dbar\bl\tr(\mu_1\cdot\psi)\wedge\omega^{n-1}\br= \int_X
d\bl\tr(\mu_1\cdot\psi)\wedge\omega^{n-1}\br=0.
$$
This proves the first part. As to the second part, a direct
computation shows that
$$0=\tr(\Dbarsta_0 \mu_2\cdot \phi)\cdot\omega^n=-2n\tr(D\pri_0\mu_2\cdot
\phi)\cdot\omega^{n-1}.
$$
The identity follows immediately.
\end{proof}

We now evaluate $\dot r(0,t)$ and $\ddot r(0,t)$. First, we show
that
\begin{equation}\lab{rone}
\dds L_s(\hh\lst)|_{s=0}=\dds \tilde L_s(\hh\lst)|_{s=0}=0.
\end{equation}
Because $F_{0,H\lzt}\wedge\omega^{n-1}=0$, the first identity
holds automatically. We now look at the second identity. By
definition, there is a function $c(s,t)$ with $c(0,t)=0$ so that
$$c(s,t)J=L_s(H\lst)=H\lst\umhalf F_{s,H\lst} H\lst\uhalf.
$$
Taking derivative of $s$ at $s=0$ gives us
$$
\dot c(0,t)J=\dds\bl H\lst\umhalf F_{s,H\lst}
H\lst\uhalf\wedge\omega^{n-1}\br|_{s=0}= H\lzt\umhalf \dot
F_{0,H\lzt} H\lzt\uhalf\wedge\omega^{n-1}.
$$
Using the explicit form of $H\lzt$, $A_0=0$, $\dot F_0=D_0\pri\dot
A_0-\Dbar_0 \dot A_0\sta$ and $D H\lzt=0$, we obtain
$$\dds\int_X\tr\bl H\lst\umhalf F_{s,H\lst}
H\lst\uhalf\wedge\omega^{n-1}\cdot I_1\br|_{s=0}=\int_X\tr\bl
(D_0\pri \varphi+ \Dbar_0\varphi\pri)\cdot I_1\br\wedge
\omega^{n-1}
$$
for some smooth sections $\varphi$ and $\varphi\pri$. The right
hand side of the above identity is zero by Lemma \ref{2.2}; thus
$$\int_X \dot c(0,t)\tr\bl J\cdot I_1\br=0,
$$
which forces $\dot c(0,t)=0$. This proves (\ref{rone}), and $\dot
r(0,t)=0$ for all $t$.

We next compute $\ddot r(0,t)$. Because of (\ref{rone}),
$$\frac{d^2}{ds^2} L_s(H\lst)|_{s=0}=
H\lzt\umhalf\frac{d^2}{ds^2}\tilde L_s(H\lst)|_{s=0} H\lzt\uhalf.
$$
A direct computation shows that
\begin{equation}\lab{2.14}
\frac{d^2}{ds^2}\tilde L_s(\hh\lst)|_{s=0} =\ddot F_0-2[\dot
A_0,[\dot A_0\sta, \hh_{0,t}]\hh_{0,t}\upmo]- 2[\dot A_0,\d_0 \dot
\hh_{0,t}\cdot \hh_{0,t}\upmo]+\Dbar_0\dot\varphi_{0,t}
\end{equation}
with
\begin{equation}\lab{2.26}
\varphi_{s,t}=\dds(\d_s \hh\lst\cdot \hh\lst\upmo) =(\dot D\pri_s
\hh_{s,t}+\d_s\dot \hh_{s,t})\hh_{s,t}\upmo =[\dot A_s\sta,
\hh_{s,t}]\hh_{s,t}\upmo+D\pri_s\dot \hh_{s,t}\cdot
\hh_{s,t}\upmo.
\end{equation}
Because $H\lzt$ commutes with $I_1$,
$$%\begin{equation}\lab{2.15}
\ddot r(0,t)={\sqrt{-1}}\left(\int_X\tr\bl\ddot
F_0\cdot\ii_1\br\wedge\omega^{n-1} -2\int_X\tr\bl[\dot A_0,[\dot
A_0\sta, \hh\lzt]\hh\lzt\upmo]\cdot
\ii_1\br\wedge\omega^{n-1}-\qquad\right.
$$%\end{equation}
$$\qquad\qquad\qquad\qquad\left.
-2\int_X \tr\bl [\dot A_0,D\pri_0\dot \hh\lzt\cdot
\hh\lzt\upmo]\cdot \ii_1\br\wedge\omega^{n-1} +\int_X \tr\bl
\Dbar_0\dot\varphi_{0,t}\cdot \ii_1\br\wedge\omega^{n-1}\right).
$$

To analyze the sign of the above integration, we use the splitting
$E=E_1\oplus E_2$ to express
$$\dot A_0=\left(%
\begin{array}{cc}
  C_{11} & C_{12} \\
  C_{21} & C_{22} \\
\end{array}%
\right).
$$
Because
$$\hh\lzt=\left(%
\begin{array}{cc}
  \exp({\frac{t}{r_2}})\cdot\ii_1 & 0 \\
  0 & \exp(\frac{-t}{r_1})\cdot\ii_2 \\
\end{array}%
\right),
$$
the second term
$$-2\sqrt{-1}\int_X\tr\bl[\dot A_0,[\dot
A_0\sta, \hh\lzt]\hh\lzt\upmo]\cdot \ii_1\br\wedge\omega^{n-1}
$$
in $\ddot r(0,t)$ is, for $\alpha=\frac{1}{n_1}+\frac{1}{n_2}$,
$$-2\sqrt{-1}(1-e^{-\alpha t})\int_X\tr\bl C_{12}\wedge C_{12}\sta\br
\wedge \omega^{n-1}-2\sqrt{-1}(1-e^{\alpha t})\int_X\tr\bl
C_{21}\sta\wedge C_{21}\br\wedge \omega^{n-1}.
$$
Similarly, because of (\ref{2.11}) and $F_s^{2,0}=F_s^{0,2}=0$,
$${\sqrt{-1}}\int_X \tr\bl \ddot F_0\cdot\ii_1\br\wedge \omega^{n-1}
=2\sqrt{-1}\int_X\tr\bl C_{12}\wedge
C_{12}\sta\br\wedge\omega^{n-1} +2\sqrt{-1}\int_X\tr\bl
C_{21}\sta\wedge C_{21}\br\wedge\omega^{n-1}.
$$
The last term in $\ddot r(0,t)$ is zero because of Lemma
\ref{2.2}; the next-to-last term is
$$-2\sqrt{-1}\int_X \tr\bl \dot A_0 \cdot\d\dot
\hh\lzt\cdot \hh\lzt\upmo\cdot\ii_1\br\wedge\omega^{n-1}+
2\sqrt{-1}\int_X\tr\bl D\pri_0 \dot \hh\lzt\cdot
\hh\lzt\upmo\cdot\dot A_0\cdot\ii_1\br\wedge\omega^{n-1},
$$
which vanishes because $\Dbarsta_0 \dot A_0=0$ and Lemma
\ref{2.2}. Therefore,
$$\ddot r(0,t)=\sqrt{-1}e^{-\alpha t}\int_X \tr(C_{12}\wedge
C_{12}\sta)\wedge\omega^{n-1}+\sqrt{-1}e^{\alpha t}\int_X
\tr(C_{21}\sta\wedge C_{21})\wedge\omega^{n-1}.
$$
Because the associated cohomology class $[C_{ij}]=\kappa_{ij}$ and
$\kappa_{21}$ and $\kappa_{12}$ are both non-zero,
$$A=\sqrt{-1}\int_X\tr(C_{12}\wedge
C_{12}\sta)\wedge\omega^{n-1} \and B=-\sqrt{-1}\int_X
\tr(C_{21}\sta\wedge C_{21})\wedge\omega^{n-1}
$$
are positive. Hence for sufficiently small $s$, the value
$r(s,t)>0$ for $t<\frac{1}{2\alpha}\ln\frac A B$ and $r(s,t)>0$
for $t>\frac{1}{2\alpha}\ln\frac A B$. Hence there is a function
$t=\rho(s)$ so that $r(s,\rho(s))=0$. This proves that the system
$L_s(H)=0$ is solvable for small $s$. Here the function $\rho(s)$
is not necessarily continuous, but $\lim_{s\to 0}\rho(s)=
\frac{1}{2\alpha}\ln\frac A B$.

%ch2.tex

\section{Linearization of Strominger's system}

In this section we will study the linearization of Strominger's
system. Before we do this, we will first rephrase the system
(\ref{0.1})-(\ref{0.4}) in the form that is easier to handle.

We fix a Calabi-Yau threefold $(X,\omega_0)$ and a
$(3,0)$-holomorphic form $\Omega$ so that
$\Omega\wedge\bar\Omega=\omega_0^3$. We let $(E,\Dbar)$ be a rank
$r$ holomorphic bundle over $X$ such that $c_1(E)=0$ and
$c_2(E)=c_2(X)$. We then choose a hermitian metric $H$ on $E$ and
let $D_\hh=D_\hh\pri\oplus \Dbar$ be the hermitian connection of
$(E,\Dbar,\hh)$; its curvature $F_\hh=D_\hh\circ D_\hh$ satisfies
$$F_\hh^{2,0}=F^{0,2}_\hh=0.
$$
Thus the second equation of the Strominger's system follows
automatically.

The fourth equation of the system is a non-linear equation of a
hermitian form $\omega$ involving the adjoint $d_\omega\sta$ of
$\omega$. It turns out that this equation is equivalent to
$$d(\|\Omega\|_{\omega}\omega^2)=0.
$$

We now prove this equivalence. We let $\cH(X)$ and $\cK(X)$ be the
cones of positive definite hermitian forms and Kahler forms on $X$
respectively. Given an $\omega\in\cH(X)$, we let $\ast_\omega$ be
the (hermitian) star operator of $\omega$; and let $d\sta_\omega$
be the adjoint of $d$ with respect to the metric $\omega$.

The hermitian star operator has an explicit local expression.
Given a hermitian form $\omega$ on $X$ it induces canonical
hermitian metrics on $T_{X,\CC}$ and on $\wedge^k T\dual_{X,\CC}$.
Let $(\cdot,\cdot)_\omega$ be the hermitian metric on $\wedge^k
T\dual_{X,\CC}$ and $\frac{1}{3!}\omega^3$ its associated volume
form on $X$. The star operator $\ast_\omega$ is the $\CC$-linear
operator defined via
$$(\varphi,\psi)_\omega\cdot\frac{\omega^3}{3!}
=\varphi\wedge\ast_\omega\bar\psi.
$$
Let $p\in X$ be any point and let $\varphi_1,\varphi_2,\varphi_2$
be an $(\cdot,\cdot)_\omega$-orthonormal basis (a moving frame) of
the $(1,0)$-forms near $p$ obeying
$(\varphi_i,\varphi_j)_\omega=2\delta_{ij}$. Then the hermitian
form
$$\omega=\sqrtt\sum_{i=1}^3 \varphi_i\wedge\bar\varphi_i.
$$
For any subset $I=\{i_1,\cdots,i_k\}\sub\{1,2,3\}$, we denote by
$\varphi_I=\varphi_{i_1}\wedge\cdots\wedge\varphi_{i_k}$, and
denote by $I\ucirc$ the complement $\{1,2,3\}-I$.
%Hence for instance $\varphi_{2\ucirc}=\varphi_1\wedge\varphi_3$.
Following this convention,
\begin{equation}\lab{3.41}
\ast_\omega(c\,\bar\varphi_I\wedge\varphi_J)=\epsilon_{IJ}\sqrto\
2^{|I|+|J|-3} c\,
\varphi_{I\ucirc}\wedge\bar\varphi_{J\ucirc},\quad c\in\CC,
\end{equation}
where $\epsilon_{IJ}$ is the parity of permuting
$(I,J;I\ucirc,J\ucirc)\mapsto (1,2,3;1\pri,2\pri,3\pri)$.

We now re-state and prove the mentioned equivalence.

\begin{lemm}
Let $\omega_0$ be the reference Kahler form as before. Then the
equation (\ref{0.4}) is equivalent to
\begin{equation}\lab{2.4}
\ast_{\omega_0} d\,(\|\Omega\|_{\omega}\omega^2)=0.
\end{equation}
\end{lemm}

\begin{proof}
Let $f$ be a positive real valued function, then
$$d(f\omega^2)=f\, d\,\omega^2+df\wedge\omega^2=2f
d\ast_\omega\omega+df\wedge\omega^2.
$$
Thus
%\begin{eqnarray*}
$$\ast_\omega d(f\omega^2)=2f\ast_\omega d\ast_\omega
\omega+\ast_\omega(df\wedge\omega^2)= -2f d_\omega\sta\omega+2d_c
f.
%\,=\,2f(-d\sta_\omega\omega-d_c\log f\upmo)\end{eqnarray*}
$$
Here we have used the identity
$$\ast_\omega(df\wedge\omega^2)=2d_cf,
$$
which holds for all hermitian form $\omega$. Replacing $f$ by
$\|\Omega\|$, we obtain
$$\ast_\omega d\bl\|\Omega\|_\omega\omega^2)=
2\|\Omega\|_\omega(-d_\omega\sta\omega+ d_c
\log\|\Omega\|_\omega\br,
$$
which vanishes if and only if
$$d_\omega\sta \omega=d_c\log \|\Omega\|_\omega.
$$
Finally, since $\ast_\omega$ and $\ast_{\omega_0}$ are both
isomorphisms, $\ast_\omega
d\bl\|\Omega\|_\omega\upmo\omega^2\br=0$ if and only if
$$\ast_{\omega_0} d\bl\|\Omega\|_\omega\omega^2\br=0.
$$
This proves the lemma.
\end{proof}

To apply the implicit function theorem, we need to specify the
range of the operators associated to Strominger's system. For
that, noting that $2dd_c=\sqrt{-1}\partial\dbar$, we let
$R(dd_c)\sub \Omega_\RR^{2,2}(X)$ and $R(d\sta_\omz)\sub
\Omega_\RR^1(X)$ be the range of
$$dd_c: \Omega_\RR^{1,1}(X)\to \Omega_\RR^{2,2}(X)\and
d_\omz\sta: \Omega_\RR^{1,1}(X)\to \Omega_\RR^1(X).
$$
Because $(X,\omz)$ is a Kahler manifold, by $\partial\dbar$-lemma,
a real form $\alpha\in R(dd_c)$ if and only if $d\alpha=0$. Hence,
after picking a usual Banach norm on $\Omega_{\RR}^{2,2}(X)$,
$R(dd_c)$ is closed in it. As to $R(d_\omz\sta)$, since
$d_\omz\sta$ is part of an elliptic complex, it is also closed.
This way, after replacing (\ref{0.4}) by (\ref{2.4}) and omitting
the equation (\ref{0.2}), the Strominger's system is equivalent to
the vanishing of the operator
\begin{equation}\lab{3.10}
\LL=\LL_1\oplus\LL_2\oplus\LL_3: \cH(E)_1\times\cH(X)\lra
\Omega_\RR^{6}(\Endo E)\oplus R(dd_c)\oplus R(d_{\omega_0}\sta),
\end{equation}
%where $\cH(E)_1$ is the space of determinant one\footnote{We say a
%hermitian metric on $E$ has determinant one if its induced metric
%on $\wedge^r E\cong\CC_X$ is the constant one metric.} hermitian
%metrics on $E$,
defined by
\begin{equation}\lab{3.1}
\LL_1(\hh,\omega)=H\umhalf  F_\hh H\uhalf\wedge \omega^2\in
\Omega_\RR^{6}(\Endo E);
\end{equation}
\begin{equation}\lab{3.2}
\LL_2(\hh,\omega)=\half dd_c\omega+\bl\tr(F_\hh\wedge
F_\hh)-\tr(R_\omega\wedge R_\omega)\br \in \Omega_\RR^{2,2}(X);
\end{equation}
\begin{equation}\lab{3.3}
\LL_3(\hh,\omega)=\ast_{\omega_0}
d\,(\|\Omega\|_{\omega}\omega^2)\in \Omega^1_\RR(X).
\end{equation}
Because $c_2(E)=c_2(TX)$ and $X$ is a Kahler manifold, by
$\partial\dbar$-lemma the image of $\LL_2$ lies in $R(P)$. As to
$\LL_3$, because
$$\ast_{\omega_0} d=\pm\ast_{\omega_0} d\ast_{\omega_0}\ast_{\omega_0}\upmo
=\mp d_{\omega_0}\sta\ast_{\omega_0}\upmo,
$$
its image lies in the range of $d_{\omega_0}\sta$ as well.
Therefore the operator $\LL$ is well-defined.

\begin{prop}
Suppose $\LL(H,\omega_0)=0$. Then the three summands of the
linearization of $\LL$ at $(H,\omega_0)$ are
\begin{eqnarray*}
\delta\LL_1(H,\omega_0)(\delta h,\delta\omega)&=&\Dbar
\d_{H}\delta h\wedge\omega_0^2+2 H\umhalf F_{H}H\uhalf \wedge\omega_0\wedge\delta\omega;\\
\delta\LL_2(H,\omega_0)(\delta h,\delta\omega)&=& \half
dd_c\delta\omega +2\bl\tr(\delta F_{H}(\delta h)\wedge F_{H})-
\tr(\delta R_{\omega_0}(\delta \omega)\wedge R_{\omega_0})\br;
\\
\delta\LL_3(H,\omega_0)(\delta
h,\delta\omega)&=&2d_{\omega_0}\sta\delta\omega-d_{\omega_0}\sta((\delta
\omega,\omega_0)_\omz\omega_0).
\end{eqnarray*}
\end{prop}

Here as before we follow the convention $\delta H=H\umhalf \delta
h H\umhalf$.

\begin{proof}
The formula for $\delta\LL_1$ is well-known \cite{YU}; the formula
for $\delta\LL_2$ in the written form is a tautology; we stop
short of finding an explicit form of $\delta R$ since the current
form is sufficient for our purposes.

We now prove the formula for $\delta\LL_3$. Let $\omega_t$ be a
smooth variation of the hermitian form $\omega_0$; let
$\varphi_1(t),\varphi_2(t),\varphi_3(t)$ be an orthonormal basis
of $(1,0)$-forms, smooth in $t$, expressed in a holomorphic
coordinate $(z_1,z_2,z_3)$ near $p\in X$ by
$$\varphi_i(t)=\sum_j b_{ij}(t) dz_j,\quad b\lij(0)(p)=\delta\lij
\and (\varphi_i(t),\varphi_j(t))_{\omega_t}=2\delta_{ij}.
$$
We can compute explicitly $\ddt(\omega_t^2)|_{t=0}$. First,
$$\omega_t^2=\frac{1}{2}\sum
\varphi_{i\ucirc}(t)\wedge\bar\varphi_{i\ucirc}(t)
=\frac{1}{2}\sum_{i,l,k}c_{ik}(t)\,\bar
c_{il}(t)\,dz_{k\ucirc}\!\wedge d\bar z_{l\ucirc},
$$
where $c_{ij}(t)$ is the $ij$-th minor of the matrix
$(b\lij(t))_{3\times 3}$; namely
\begin{equation}\lab{2.6}
\bl c_{ij}(t)\br^t=\det(b\lij(t))\cdot\bl b\lij(t)\br\upmo.
\end{equation}
Hence at $p$,
$$\ddt\omega_t^2|_{t=0}=\frac{1}{2}
\sum\bl \dot c_{lk}(0)+\dot{\bar c}_{kl}(0)\br
dz_{k\ucirc}\!\wedge d\bar z_{l\ucirc}.
$$
Using the identity (\ref{2.6}) above,
$$\dot c_{lk}(0)+\dot{\bar c}_{kl}(0)=-\dot
b_{kl}(0)-\dot{\bar b}_{lk}(0)+c_{lk}(0)\sum\dot b_{ii}(0)+\bar
c_{kl}(0)\sum_i \dot{\bar b}_{ii}(0).
$$
Therefore at $p$,
$$\ddt\omega_t^2|_{t=0}=\frac{-1}{2}\sum_{l,k} (\dot
b_{kl}(0)+\dot{\bar b}_{lk}(0))dz_{k\ucirc}\!\wedge d\bar
z_{l\ucirc} +\frac{1}{2}\bl\sum_k dz_{k\ucirc}\!\wedge d\bar
z_{k\ucirc}\br \cdot \bl\sum \dot b_{ii}(0)+\dot{\bar
b}_{ii}(0)\br.
$$
%Since $b_{ij}$ is hermitian symmetric,
%For simplicity, we denote $\dot b_{ii}(0)=\dot{\bar b}_{ii}(0)$.
On the other hand, $\omega_0^2=\frac{1}{2}\sum dz_{k\ucirc}\wedge
d\bar z_{k\ucirc}$. Hence
\begin{equation}\lab{4.4}
\ddt\omega_t^2|_{t=0}=\frac{-1}{2}\sum_{l,k} (\dot
b_{kl}(0)+\dot{\bar b}_{lk}(0))dz_{k\ucirc}\!\wedge d\bar
z_{l\ucirc} +\omega_0^2\bl\sum \dot b_{ii}(0)+\dot{\bar
b}_{ii}(0)\br.
\end{equation}
Next we compute
$$\ddt\log\|\Omega\|^2_{\omega_t}|_{t=0}=-\ddt
\frac{\omega_t^3}{\omega_0^3}|_{t=0}=-\sum \dot
b_{ii}(0)+\dot{\bar b}_{ii}(0).
$$
Adding $\|\Omega\|_{\omega_0}\equiv1$, we get
\begin{eqnarray*}
\ddt\bl\|\Omega\|_{\omega_t}\omega_t^2\br|_{t=0}&=&\left(
\half\omega_0^2\ddt\log\|\Omega\|^2_{\omega_t}+\ddt\omega_t^2\right)|_{t=0}\\
&=&-\frac{1}{2}\sum_{l,k}(\dot b_{kl}(0)+\dot
b_{lk}(0))dz_{k\ucirc}\wedge d\bar z_{l\ucirc}+ \half\bl\sum \dot
b_{ii}(0)+\dot{\bar b}_{ii}(0)\br\omega^2_0.
\end{eqnarray*}
On the other hand, at $p$
$$\ddt\omega_t|_{t=0}=\sqrtt\sum_i \dot\varphi_i\wedge{\bar\varphi}_i+
\varphi_i\wedge\dot{\bar\varphi}_i = \sqrtt\sum_{i,j}(\dot
b_{ji}(0)+\dot{\bar b}_{ij}(0))dz_i\wedge d\bar z_j.
$$
Hence
$$\ast_{\omega_0}\dot \omega_0=\frac{1}{4}\sum_{i,j}(\dot{\bar b}_{ji}(0)+\dot
b_{ij}(0))dz_{i\ucirc}\!\wedge d\bar z_{j\ucirc}.
$$
Combined, we obtain
$$\ddt\bl\|\Omega\|_{\omega_t}\omega_t^2\br|_{t=0}=
-2\ast_{\omega_0}\dot\omega_0+\bl\sum \dot b_{ii}(0)+\dot {\bar
b}_{ii}(0)\br\frac{\omega_0^2}{2}.
$$
It remains to treat the term $\sum\dot b_{ii}(0)+\dot {\bar
b}_{ii}(0)$. From
$$(\dot\omega_0,\omega_0)_{\omega_0}\frac{\omega^3_0}{3!}
=\dot\omega_0\wedge\ast_{\omega_0}\omega_0
\and\ast_{\omega_0}\omega_0=\frac{1}{4}\sum dz_{k^\circ}\wedge
d\overline{z}_{k^\circ},
$$
we get
\begin{eqnarray*}
\dot\omega_0\wedge\ast_{\omega_0}\omega_0
&=&\frac{\sqrt{-1}}{8}\sum(\dot{b}_{ij}(0)+\dot{\overline{b}}_{ji}(0))dz_i\wedge
d\overline{z}_j\wedge dz_{k^\circ}\wedge d\overline{z}_{k^\circ}\\
&=&-\frac{\sqrt{-1}}{8}\sum(\dot{b}_{ii}(0)+\dot{\overline{b}}_{ii}(0))dz_1\wedge
d\overline{z}_1\wedge dz_2\wedge d\overline{z}_2\wedge dz_3\wedge
d\overline{z}_3;
\end{eqnarray*}
hence
$$(\dot{\omega}_0,\omega_0)_{\omega_0}=
\frac{\dot\omega_0\wedge
\ast_{\omega_0}\omega_0}{{\omega^3_0}/{3!}}=\sum\dot{b}_{ii}(0)+\dot{\bar
b}_{ii}(0).
$$
This proves that
$$\ddt\bl\|\Omega\|_{\omega_t}\omega_t^2\br|_{t=0}=
-2\ast_{\omega_0}\dot\omega_0+\bl\sum\dot{b}_{ii}(0)+\dot{\bar
b}_{ii}(0)\br\omega_0^2=
-2\ast_{\omega_0}\dot\omega_0+\ast_{\omega_0}
(\dot{\omega}_0,\omega_0)_{\omega_0}\omega_0.
$$
Finally,  Applying $\ast_{\omega_0} d$ to both sides of this
identity, we obtain
$$\ddt\ast_{\omega_0}
d\bl\|\Omega\|_{\omega_t}\omega_t^2\br|_{t=0}=2d\sta_{\omega_0}\dot\omega_0
-d\sta_{\omega_0}((\dot{\omega}_0,\omega_0)_{\omega_0}\omega_0).
$$
This proves the Proposition.
\end{proof}

Strominger's system admits a class of reducible solutions. Let
$$(E,\Dbar_0)=\CC_X\ur\oplus TX
$$
be the direct sum of the trivial holomorphic bundle $\CC_X\ur$ and
the tangent bundle $TX$. We fix an isomorphism $\wedge^{r+3}
E\cong\CC_X$; we endow $E$ with the hermitian metric $<\,,>$ that
is a direct sum of a constant metric on $\CC_X\ur$ and the
Calabi-Yau metric $\omega_0$ on $TX$. We normalize $<\,,>$ so that
its induced metric on $\wedge^{r+3} E\cong\CC_X$ is the constant
one metric. As before, the metric $<\,,>$ is identified with the
identity endomorphism $I\mh E\to E$.

Now let $\cH_{r\times r}^+$ be the space of positive definite
hermitian symmetric $r\times r$ metrics; let $I_1$ and $I_2$ be
the identity endomorphisms of $\CC_X\ur$ and $TX$ respectively. By
abuse of notation, for $T\in \cH^+_{r\times r}$ we also view it as
the constant endomorphism of $\CC_X\ur$ induced by $T$, viewed as
an endomorphism of $E$. Then the assignment
$$T\in\cH_{r\times r}^+ \longmapsto H_T=T\oplus
|T|^{-1/3}I_2\in \cH(E)_1,\ |T|=\det T,
$$
associates each $T\in\cH_{r\times r}^+$ to a hermitian metric of
$E$.

Obviously, the hermitian curvature $F_{H_T}$ of $(E,<\,,>_{H_T})$
is $0\oplus R_{\omega_0}$; hence $F_{H_T}\wedge
F_{H_T}=R_{\omega_0}\wedge R_{\omega_0}$. Because $\omega_0$ is
$d$-closed,
$$\LL_2({H_T},\omega_0)=\half dd_c\omega_0+
\tr(F_{{H_T}}\wedge F_{H_T})-\tr(R_{\omega_0}\wedge
R_{\omega_0})=0.
$$
Further, because $<\,,>_{H_T}$ is Hermitian-Yang-Mills, and
because $d_{\omega_0}\sta\omega_0=0$ and
$\Omega\wedge\bar\Omega=\omega_0^3$,
$\LL_1({H_T},\omega_0)=\LL_3({H_T},\omega_0)=0$. Therefore
$({H_T},\omega_0)$ is a solution to $\LL(\hh,\omega)=0$. Indeed,
for any constant $c>0$, the pair $({H_T},c\omega_0)$ is a solution
to $\LL=0$. These solutions are reducible because the vector
bundle $E$ splits under the hermitian connection $D_{H_T}$. In
this paper, we will call such solutions the {\sl trivial}
solutions to Strominger's system.

To construct irreducible solutions to Strominger's system, we will
first form a family of holomorphic structures $\Dbar_s$ on $E$
that is a smooth deformation of $\Dbar_0$; we then show that
certain trivial solutions to Strominger's system on $(E,\Dbar_0)$
can be extended to (irreducible) solutions on $(E,\Dbar_s)$. We
shall prove this by applying implicit function theorem to the
operator $\bL$ of (\ref{3.10}).

To this end, we pick an integer $k$ and a large $p$ and endow the
domain and the target of $\LL$ the Banach space structures as
indicated:
$$\cH(E)_{1,L^p_k}\times\cH(X)_\lpk\lra
\Omega_\RR^{6}(\Endo E)_\lpkt\oplus R(dd_c)_\lpkt\oplus R(
d\sta_{\omega_0})_\lpko.
$$
$\bL$ becomes a smooth operator and its linearized operator
$\delta\LL$ at a solution $(H,\omega)$ becomes a linear map
$$\Omega^0(\her E)_\lpk\oplus\Omega^{1,1}_\RR(X)_\lpk\lra
\Omega_\RR^{6}(\Endo E)_\lpkt \oplus R(dd_c)_\lpkt\oplus R(
d\sta_{\omega_0})_\lpko.
$$
Here we used the canonical isomorphisms $T_H\cH(E)_{1,\lpk}\cong
\Omega^0(\her E)_\lpk$ and $T_{\omega}\cH(X)_\lpk\cong
\Omega^{1,1}_\RR(X)_\lpk$. For simplicity, in the following we
will abbreviate
$$\cW_1=\Omega_\RR^{6}(\Endo E)_\lpkt
\and \cW_2=R(dd_c)_\lpkt\oplus R( d\sta_{\omega_0})_\lpko.
$$
Thus $\delta\LL(H,\omega)$ is a linear map
\begin{equation}
\lab{3.19} \delta\LL(H,\omega): \Omega^0(\her
E)_\lpk\oplus\Omega^{1,1}_\RR(X)_\lpk\lra \cW_1\oplus\cW_2.
\end{equation}

To study the kernel and the cokernel of $\delta\LL$ at a trivial
solution $(H_T,c\omega_0)$ we will first look at the linear map
% $\FF_\hh=\omega_0^2\wedge F_\hh$:
\begin{equation}\lab{3.11}
\FF(\delta h)= \Dbar_0 D\pri_{0,H_T}(\delta h)\wedge\omega_0^2 :
\Omega^0(\her E)_\lpk\lra \Omega^{6}_\RR(\Endo E)_{\lpkt}.
\end{equation}
Here according to our convention, $D_{H_T}=D_{0,H_T}\pri\oplus
\Dbar_{0}$ is the hermitian connection of $(E,\Dbar_0,H_T)$ for a
$T\in \cH_{r\times r}^+$. Since $(E,\Dbar_0)=\CC_X\ur\oplus TX$,
the above is a linear elliptic operator of index $0$ whose kernel
is
$$V_0=\{M\oplus a I_2\mid M\in\End( \CC\ur), M=M\sta,\
\tr M+3a=0\}
$$
and cokernel is
\begin{equation}\lab{3.21}
V_1=\omega_0^3\cdot V_0 \sub \cW_1=\Omega^{6}_\RR(\Endo
E)_{\lpkt}.
\end{equation}
We let $\bP$ be the obvious projection
$$\bP: \cW_1\lra\cW_1/V_1.
$$

\begin{prop}\lab{p3.3}
Let $(X,\omega_0)$, $\Omega$, $<\,,>$ and $T\in\cH_{r\times r}^+$
be as before. Then there is a constant $C$ so that for any $c>C$,
the linear operator
$$\bP\circ\delta\cL_1(H_T,c\omega_0) \oplus\delta\cL_2(H_T,c\omega_0)
\oplus \delta\cL_3(H_T,c\omega_0)
$$
from $\Omega^0(\her E)_{\lpk} \oplus\Omega_\RR^{1,1}(X)_\lpk$ to
$\cW_1/V_1\oplus\cW_2$ is surjective.
\end{prop}

\begin{proof}
As we shall see, the proof of the Proposition relies on the
understanding of the operator
$$T:\Omega_\RR^{1,1}(X)_\lpk\lra \cW_2
$$
defined by, after setting $P=\half d d_c=\sqrt{-1}\partial\dbar$,
$$T\mu=
\bl P\mu,2d_{\omega_0}\sta\mu
-d_{\omega_0}\sta\bl(\mu,\omega_0)_{\omega_0}\omega_0\br\br.
$$
Before we go on, we remark that since in the proof of this
Proposition we will solely work with the Kahler form $\omega_0$,
for convenience we will abbreviate $\ast_\omz$ and $d\sta_\omz$ to
$\ast$ and $d\sta$.

For the starter, we form the linear operator $S$:
$$S\mu=2\mu-(\mu,\omz)_{\omz}\omz:
\Omega_{\RR}^{1,1}\lra\Omega_{\RR}^{1,1}
$$
and its inverse
$$S\upmo\phi=\half(\phi-(\phi,\omz)_\omz\omz).
$$
Then by setting $\phi=S\mu$, $T\mu$ can be expressed as
$$ T\mu=T\circ S\upmo\phi
=\left( P\circ S\upmo\phi, d\sta\phi\right).
$$
Then applying the Hodge decomposition to
$\phi\in\Omega_\RR^{1,1}(X)$,
$$\phi=dd\sta\psi+d\sta d\psi+h
$$
for a real $(1,1)$-form $\psi$ and harmonic $h$. By the
$\partial\dbar$-lemma, we can rewrite $d\sta d\psi=\ast P\alpha$
for a real form $\alpha$.

As to the harmonic $h$, we check that the pairing $(h,\omz)_\omz$
is constant. Since $(X,\omz)$ is Kahler,
$$d_c \ast h=
d\sta \ast h\wedge\omz-d\sta(\ast h\wedge\omz);
$$
and since $d\sta\ast h=d_c \ast h=0$, $d\sta(\ast h\wedge\omz)=0$.
Hence the defining identity
\begin{equation}\lab{51}(h,\omz)\!\ast\!1
=\ast h\wedge\omz
\end{equation}
forces $(h,\omz)_\omz$ to be a constant. Therefore the space of
harmonic forms $\HH\sub \Omega_\RR^{1,1}(X)$ lies in the kernel of
both $T$ and $T\circ S\upmo$.

With this said, to study the surjectivity of $T$ we only need to
look at those $\phi$ that are orthogonal to $\HH$ under the
$L^2$-intersection pairing
$$<\!u,v\!>=\int_X(u,v)_\omz\ast 1.
$$
In particular, such $\phi$ has decomposition
$$\phi=\ast P\alpha+d\sta d\psi,
$$
and
$$T\circ S\upmo\phi=\left(P\circ S\upmo(\ast P\alpha)+P\circ
S\upmo(dd\sta \psi), d\sta d(d\sta\psi)\right).
$$

To proceed, we look at the operator $U$:
\begin{equation}\lab{31}
U\alpha=2 \ast\! P \circ S\upmo(\ast P\alpha) =\ast P\bl\ast
P\alpha- (\ast P\alpha,\omz)_\omz \omz\br.
\end{equation}
Because
$$ P\sta=\bl\sqrt{-1}\partial\dbar\br\sta=
-\sqrt{-1}{\dbar}\sta\partial\sta=\ast\sqrt{-1}\partial\dbar\ast
=\ast P\ast,
$$
$U\alpha$ can be re-written as
\begin{equation}\lab{34}
U\alpha= P\sta P\alpha-\ast P\bl(\ast
P(\alpha),\omz)_{\omz}\omz\br.
\end{equation}

To proceed, we need to simplify the operator $U$. We first use the
identities
\begin{equation}\lab{33}
\partial\sta\mu\wedge\omz-\partial\sta(\mu\wedge\omz)=
\sqrt{-1}\dbar\mu \and
\dbar\sta\mu\wedge\omz-\dbar\sta(\mu\wedge\omz)=-\sqrt{-1}\partial\mu,
\end{equation}
which hold for all Kahler manifolds, to derive
$$\partial\sta(f\omz^2)=-2\sqrt{-1}\dbar f\wedge\omz.
$$
Using $(\ast P\alpha,\omz)_\omz=\ast(P\alpha\wedge\omz)$, which
follows from (\ref{51}), we have
$$P\bl(\ast P\alpha,\omz)_\omz\omz\br
=P\bl\ast(P\alpha\wedge\omz)\wedge\omz\br
=-\sqrt{-1}\ast{\dbar}\sta\partial\sta
(P\alpha\wedge\omz)\wedge\omz.
$$
Applying the identities (\ref{33}) further, we obtain
$$\partial\sta(P\alpha\wedge\omz)=
\partial\sta P\alpha\wedge\omega-
\sqrt{-1}\dbar P\alpha=\partial\sta P\alpha\wedge\omega
$$
and
$${\dbar}\sta\partial\sta \bl P\alpha\wedge\omz\br=
{\dbar}\sta\bl\partial\sta P\alpha\wedge\omz\br=
{\dbar}\sta\partial\sta P\alpha\wedge\omz+
\sqrt{-1}\partial\partial\sta P\alpha\wedge\omz.
$$
Put together, we obtain
\begin{eqnarray*} P\bl(\ast
P\alpha,\omz)_\omz\omz\br& =& -\sqrt{-1}\ast \dbar\sta\partial\sta
\bl P\alpha\wedge\omz\br\wedge\omz\\
&=&- \sqrt{-1}\ast\left( \dbar\sta\partial\sta
P\alpha\wedge\omz+\sqrt{-1}\partial\partial\sta
P\alpha\wedge\omz\right)\wedge\omz\\
&=&\ast( P\sta
P\alpha\wedge\omz)\wedge\omz+\ast\bl\partial\partial\sta
 P\alpha\wedge\omz\br\wedge\omz.
\end{eqnarray*}
Because $\partial\partial\sta P\alpha= \Box_\partial P\alpha$
since $\partial P\alpha=0$, the operator $U$ becomes
\begin{equation}\lab{54}
U(\alpha)= P\sta P\alpha-\ast\bl\ast( P\sta
P\alpha\wedge\omega)\wedge\omz\br -\ast\left(\ast\bl\Box_\partial
P\alpha\wedge\omz\br\wedge\omz\right).
\end{equation}

To continue, recall that for $\nu\in\Omega_\RR^{1,1}(X)$ such that
$(\nu,\omz)_\omz=0$, $\ast(\nu\wedge \omega_0)=-\nu$. Hence
$$
\ast(\nu\wedge\omega_0)=\ast((\nu-\frac{1}{3}(\nu,\omz)_\omz\omz)\wedge\omz)
+\frac{1}{3}\ast((\nu,\omz)_\omz\omz^2) =-\nu+(\nu,\omz)_\omz\omz
$$
and
$$
\ast(\ast(\nu\wedge\omega_0)\wedge\omz)
=\ast\left(-\nu\wedge\omz+(\nu,\omz)_\omz\ast\omz^2\right)
=\mu+(\nu,\omz)_\omz\omz.
$$
Therefore by (\ref{54}),
\begin{eqnarray*}
U\alpha&=& P\sta P\alpha-\bl P\sta P\alpha+( P\sta
P\alpha,\omz)_\omz\omz\br -\ast\bl\ast\Box_\partial
P\alpha\wedge\omz\br\\
&=&-\ast\bl\ast\Box_\partial P\alpha\wedge\omz)-( P\sta P\alpha,
\omz)_\omz\omz.
%\\&=&-\ast\bl\ast\bl\Box_\partial P\alpha+ \frac{1}{4}(P\sta
%P\alpha, \omz)_\omz\omz^2\br\wedge\omz\br.
\end{eqnarray*}

Now we are ready to derive the estimate that for a universal
constant $C$ (in the sense that it only depends on
$(X,\omega_0)$),
\begin{equation}\lab{est}
C\upmo\|T\circ S\upmo \phi\|\leq \|
P\alpha\|_\lpk+\|dd\sta\psi\|_\lpk\leq C\|T\circ S\upmo \phi\|,
\quad \forall\phi\perp\HH.
\end{equation}

First note that the first inequality holds because $T\circ S\upmo$
is a bounded operator. As to the second, because $d\sta
d(d\sta\psi)=\Box_\partial d\sta\psi$ and that $d\sta\psi$ is
orthogonal to the harmonic forms, the elliptic estimate ensures
that for a universal constant $C_1$,
$$%C_1\upmo\|\Box_\partial d\sta\psi\|_\lpkt\leq
\| d\sta\psi\|_{L^p_{k+1}}\leq C_1 \|\Box_\partial
d\sta\psi\|_\lpko \leq C_1 \|T\circ S\upmo \phi\|.
$$
Then because
$$P\circ S\upmo(dd\sta\psi)=-\half P(dd\sta\psi,\omz)_\omz\omz
$$
and because the right hand side involves the third differentiation
of $d\sta\psi$,
$$\|P\circ S\upmo(dd\sta\psi)\|_{L^p_{k-2}}\leq C_2 \|
d\sta\psi\|_{L^p_{k+1}} \leq C_1C_2 \|T\circ S\upmo \phi\|
$$
holds for a universal constant $C_2$. On the other hand,
\begin{equation}\lab{U}
\half\ast U\alpha=T\circ S\upmo\phi+\half P\circ S\upmo
(dd\sta\psi,\omz)-d\sta d(d\sta\psi),
\end{equation}
the previous estimates ensure that there is a universal constant
$C_3$ so that
\begin{equation}\lab{estu}
\| U\alpha\|_\lpkt\leq C_3 \|T\circ
S\upmo\phi\|.
\end{equation}
Because
$$d\ast(\ast\Box_\partial P\alpha\wedge\omz)=0,
$$
the formula of $U\alpha$ before (\ref{est}) gives
\begin{equation}\lab{14.3}
d(P\sta P\alpha,\omz)_\omz\wedge\omz=d (U\alpha).
\end{equation}
Combined with
$$\int_X (P\sta P\alpha,\omz)_\omz\ast 1=
\int_X ( P\alpha,P\omz)_\omz\ast 1=0,
$$
and that wedging $\omega$ forms an isomorphism from
$\Omega_\RR^{1,1}(X)$ to $\Omega_\RR^{2,2}(X)$ whose inverse has
bounded norm, (\ref{14.3}) and (\ref{estu}) implies that
$$\|(P\sta P\alpha,\omz)_\omz\|_\lpkt\leq C_4 \|U\alpha\|_\lpkt\leq
C_3C_4 \|T\circ S\upmo\phi\|.
$$
Thus for a universal constant $C_5$,
$$\|\Box_\partial P\alpha\|_\lpkt\leq C_5 \|T\circ S\upmo\phi\|.
$$
Finally, because $\Box_\partial$ is elliptic,
$$\|P\alpha\|_\lpk\leq C_6 \|T\circ S\upmo\phi\|.
$$
This proves that the second inequality in (\ref{est}) holds for a
universal constant $C$.

It remains to show that $T\circ S\upmo$ is surjective. Because
$d\sta$ surjects onto $R(d\sta)$, we only need to verify that
restricting to $\ker d\sta\cap\Omega_\RR^{1,1}(X)_\lpk$ the
operator $T\circ S\upmo$ surjects onto $R(dd_c)_\lpkt$. Because
$$R(\ast dd_c)_\lpk\sub \ker
d\sta\cap\Omega_\RR^{1,1}(X)_\lpk,
$$
it suffices to show that
\begin{equation}\lab{3.22}
\ast T\circ S\upmo(\ast P(\cdot))=\half U(\cdot): R(\ast
dd_c)_\lpk\lra R(\ast dd_c)_{L_{k-4}^p}
\end{equation}
is surjective. For this we note that the estimates derived so far
show that (\ref{3.22}) is injective and has closed range. Hence if
we can show that it is self-adjoint, it must be surjective as
well. We now show that $U$ is self-adjoint. Obviously, the first
term $P\sta P$ appeared in $U$ in (\ref{34}) is self-adjoint. As
to the second term, we observe that the $L^2$-intersection
$$%\begin{eqnarray}
<\! \ast P\bl(\ast P\alpha,\omz)_{\omz}\omz\br,\beta\!> =<\!(\ast
P\alpha,\omz)_{\omz}\ast\omz, P\beta\!> =\int_X(\ast
P\alpha,\omz)_\omz (\ast P\beta,\omz)_\omz\ast \!1.
$$
Because both $\alpha$ and $\beta$ are real, the above expression
is symmetrical in $\alpha$ and $\beta$. Therefore the operator $U$
is self-adjoint, and hence is surjective.

We are ready to prove the Proposition now. By a change of
trivialization of $\CC_X\ur$, we can assume without lose of
generality that $T=I_{r\times r}$; thus $H_T=I$. We next let $\her
E$ be the $\RR$-sub-vector bundle of $\End E$ consisting of
traceless pointwise $<\,,>$-hermitian symmetric endomorphisms of
$E$. Clearly, $T_I\cH(E)_{1,\lpk}=\Omega^0(\her E)_\lpk$. We now
define linear operators
$$\T_1,\ \T_2:
%T_I\cH(E)_{1,\lpk}\oplus\Omega_\RR^{1,1}(X)_\lpk=
\Omega(\her E)_\lpk\oplus \Omega_\RR^{1,1}(X)_\lpk\lra \cW_2
$$
that are
$$\T_1(\delta h,\delta\omega)=
\bl P\delta\omega,2d_{\omega_0}\sta\delta\omega
-d_{\omega_0}\sta\bl(\delta\omega,\omega_0)_{\omega_0}\omega_0\br\br
$$
and
$$\T_2(\delta h,\delta\omega)=2\tr(\delta F_{I}(\delta
h)\wedge F_{I})-2\tr(\delta R_{\omega_0}(\delta g)\wedge
R_{\omega_0}).
$$
Because
\begin{eqnarray*}
\delta\L_1({I},c\omega_0)(\delta h,c\delta\omega)
&=&c^2\delta\L_1({I},\omega_0)(\delta h,\delta\omega);\\
\delta\L_2({I},c\omega_0)(\delta
h,c\delta\omega)&=&\sqrto\partial\dbar c\delta\omega+2\tr(\delta
F_{I}(\delta h)\wedge F_{I})-2\tr(\delta
R_{c\omega_0}(c\delta \omega)\wedge R_{c\omega_0}) \\
&=&cP \delta\omega+2\tr(\delta F_{I}(\delta h)\wedge
F_{I})-2\tr(\delta R_{\omega_0}(\delta \omega)\wedge
R_{\omega_0}),
\end{eqnarray*}
and
\begin{eqnarray*}
\delta\L_3({I},c\omega_0)(\delta h,c\delta\omega) &=&2d\sta
c\delta\omega-d\sta \bl(c\delta\omega,\omz)_\omz\omz\br
\qquad\qquad\qquad\quad\qquad\qquad\qquad\ \
\end{eqnarray*}
\begin{equation}\lab{3.91}
\bP\circ\delta\L_1({I},c\omega_0)\oplus\delta\L_2({I},c\omega_0)
\oplus\delta\L_3({I},c\omega_0)=
c^2\bP\circ\delta\L_1({I},\omega_0)\oplus c(\T_1+c\upmo\T_2)
\end{equation}
Hence to prove the Proposition we need to show that the right hand
side is surjective. Based on the discussion before,
$$\bP\circ \delta\L_1({I},\omega_0)(\delta h, 0)=\bP\circ\FF(\delta h):
\Omega^0(\her E)_\lpk\lra \cW_1/V_1
$$
is surjective and its kernel is $V_0$. Also, we proved that
$$\T_1: \Omega_\RR^{1,1}(X)_\lpk\lra \cW_2,
$$
which is the operator $T$ discussed before, is surjective with
kernel $\HH\sub \Omega_\RR^{1,1}(X)$.

Now let $\cV\sub \Omega^0(\her E)_\lpk\oplus \Omegaoo_\lpk$ be the
orthogonal complement of $V_1\oplus \HH$. For simplicity, we
abbreviate $\T_0=\bP\circ\delta\LL_1(I,\omega_0)$. The discussion
before shows that
$$(\T_0\oplus\T_1)|_{\cV}: \cV\lra \cW_1/V_1\oplus\cW_2
$$
is surjective and that there is a constant $C$ so that
\begin{equation}\lab{3.16}
C\upmo \|(u_1,u_2)\|\leq \|\bl\T_0(u_1,u_2),\T_1(u_1,u_2)\br\|\leq
C \|(u_1,u_2)\|, \qquad (u_1,u_2)\in \cV.
\end{equation}
Because $\T_2$ is a bounded operator, for sufficiently large $c$,
$$\T_0\oplus (\T_1+c\upmo \T_2): \tmetric_\lpk\times\Omegaoo
\lra\cW_1/V_1\oplus\cW_2
$$
is surjective. In particular, the left hand side of (\ref{3.91})
is surjective. This proves the Proposition.
\end{proof}

%ch3.tex

\section{Irreducible solutions to Strominger's system}

In section two, assuming the existence of a non-degenerate
deformation of holomorphic structures of the vector bundle
$E_1\oplus E_2$ we showed how to use perturbation method to prove
the existence of the Hermitian-Yang-Mills connections. In this
section, we will construct solutions to Strominger's system using
similar method. We will find an initial trivial solution to the
Strominger's system and show that it can be extended to a family
of irreducible solutions.

We continue to work with a Calabi-Yau threefold $(X,\omega_0)$ and
the vector bundle
$$(E,\Dbar_0)=\CC_X\ur\oplus TX;
$$
we fix a smooth isomorphism
%\begin{equation}\lab{4.1}
$\mathop{\wedge}^{r+3} E\cong\CC_X$
%\end{equation}
so that the $\Dbar_0$ induces the standard holomorphic structure
on $\CC_X$; we let $\Dbar_s$ be a smooth deformation of the
holomorphic structure $\Dbar_0$. As in section two, we write
$$\Dbar_s=\Dbar_0+A_s, \qquad A_s\in \Omega^{0,1}(\End E)
$$
and write
$$\dot A_0=\left(%
\begin{array}{cc}
  C_{11} &C_{12} \\
  C_{21} &C_{22} \\
\end{array}%
\right) \in\Omega^{0,1}(\End E)
$$
according to the decomposition $E=\CC_X\ur\oplus TX$. Because of
Lemma \ref{1.31}, we can assume without lose of generality that
$C_{ij}$ are $\Dbar_0$-harmonic. Since $E_1=\CC_x\ur$ and
$H^1_{\dbar}(X,\CC_X)=0$,
\begin{equation}\lab{4.9}
C_{11}=0.
\end{equation}
Because $\Pic X$ is discrete, we can assume further that $\tr
A_s=0$ for all $s$. This means that under given smooth isomorphism
$\wedge^{r+3} E\cong\CC_X$, the induced holomorphic structure on
$\wedge^{r+3} E$ is the standard holomorphic structure on $\CC_X$.

Next, we let $H_1$ be the standard constant metric on $\CC_X\ur$
and let $H_2$ be induced by the Calabi-Yau metric $\omega_0$
normalized so that $\det(H_1\oplus H_2)$ is the constant one
metric on $\wedge^{r+3} E\cong\CC_X$. The pair of $<\,,>=H_1\oplus
H_2$ and $\omega_0$ is a trivial solution of the Strominger's
system on $(E,\Dbar_0)$. We fix such $<\,,>$ as a reference
hermitian metric on $E$. Following the convention in the previous
section, all other determinant one hermitian metrics on $E$ are of
the forms $<\cdot,\cdot>_H=<\cdot\,H,\cdot>$ for some determinant
one pointwise positive definite $<\,,>$-hermitian symmetric
endomorphisms of $E$.

Following this convention, the space of all trivial solutions to
Strominger's system on $(E,\Dbar_0)$ with Kahler form $\omega_0$
is isomorphic to the space of determinant one positive definite
$r\times r$ hermitian symmetric matrices $T$ with the
correspondence
$$T\in\cH_{r\times r}^+ \longmapsto H_T=T\oplus
|T|^{-1/3}I_2\in \cH(E)_1.
$$

With the chosen Kahler form $\omega_0$ and a hermitian metric
$H_T$, the proposition \ref{p3.3} says that for $V_1$ the cokernel
defined in (\ref{3.21}) and for large enough $c$, the linearized
operator $\delta\LL$ at $(H_T,c\omega_0)$ surjects onto
\begin{equation}\lab{4.3}
\Omega^6_\RR(\su E)_\lpkt/V_1\oplus R(dd_c)_\lpkt\oplus R(
d_{\omega_0}\sta)_\lpko.
\end{equation}

With the connection forms $A_s$, the metric $<\,,>$ and the Kahler
form $\omega_0$ so chosen, we can now define operators
$$\LL_s=\LL_{s,1}\oplus \LL_{s,2}\oplus\LL_{s,3}
$$
between
$$ \cH(E)_{1,\lpk}\times\cH(X)_\lpk\lra \Omega^6_{\RR}(\su
E)_\lpkt\oplus R(dd_c)_\lpkt\oplus R(d_{\omega_0}\sta)_\lpko
$$
with $\LL_{s,i}$ defined as in (\ref{3.1})-(\ref{3.3}) of which
the curvature form $F_H$ is replaced by the hermitian curvature of
$(E,\Dbar_s,H)$:
$$F_{s,H}=D_{s,H}\circ D_{s,H}.
$$

Let $\bP$ be the projection from
$$\Omegath(\su E)_\lpkt\oplus R(dd_c)_\lpkt
\oplus R(d_{\omega_0}\sta)_\lpko
$$
to (\ref{4.3}) and let $\cH_{\omega}(X)_\lpk$ be the space of
those $\lpk$-hermitian forms whose $\omega_0$-harmonic parts are
$\omega$.

\begin{lemm}\lab{6.3}
For any $T_0\in \cH^+_{r\times r}$, there are constants $a>0$ and
$C>0$ such that for any $c>C$ there is a neighborhood $\cU_c$ of
$(H_{T_0},c\omega_0)\in
\cH(E)_{1,\lpk}\times\cH_{c\omega_0}(X)_\lpk$ such that for each
$s\in [0,a)$ the set $\cS_s=\bl \bP\circ \bL_s\br\upmo(0)\cap
\cU_c$ is a smooth $r^2$-dimensional manifold and that the union
\begin{equation}\lab{4.11}
\cS=\coprod_{s\in[0,a)}\cS_s\times s\sub\cU_c\times[0,a)
\end{equation}
is a smooth $(r^2+1)$-dimensional manifold.
\end{lemm}

\begin{proof}
By proposition \ref{p3.3}, there is a $C>0$ such that the
linearized operator of $\bP\circ\bL_0$ is surjective at
$(H_{T_0},c\omega_0)$. Hence by the implicit theorem, for
sufficiently small $s$ the solution set to $\bP\circ\LL_s=0$ is
smooth near $(H_{T_0},c\omega_0)$ and has dimension equal to the
index of the linear operator $\bP\circ\delta\bL_0$, which is
$r^2+\dim H^{1,1}(X,\RR)$. By restricting to the slice
$$\cH(E)_{1,\lpk}\times\cH_{c\omega_0}(X)_\lpk\sub
\cH(E)_{1,\lpk}\times\cH(X)_\lpk
$$
that is transversal to the kernel of $\bP\circ\delta\LL_0$, the
solution set $\cS_s$ will have the property as stated in the
Lemma. This proves the Lemma.
\end{proof}

Following our convention, $\cS_0$ consists of all pairs
\begin{equation}\lab{4.10}
(H\lzT,\omega\lzT); \quad H\lzT=T\oplus |T|^{-1/3}I_2, \quad
\omega\lzT=c\,\omega_0.
\end{equation}
Since $\cS_s$ and $\cS$ are smooth, by shrinking $\cU_c$ if
necessary, we can parameterize $\cS$ smoothly by $(s,T)$ so that
$(s,T)$ parameterizes the set $\cS$ that is consistent with the
projection $\cS\to [0,a)$ and the parameterization (\ref{4.10}).
By shrinking $\cU_c$ if necessary, we can assume that under this
parameterization, $\cS\cong [0,a)\times B_\epsilon(T_0)$, where
$B_\epsilon(T_0)$ is the ball of radius $\epsilon$ centered at
$T_0$ in $\cH^+_{r\times r}$. In the following, we denote by
$$(H\lsT,\omega\lsT)\in \cS_s, \quad T\in B_\epsilon(T_0),
$$
the solutions with parameters $(s,T)$. For simplicity, we denote
by $F\lsT$ the curvature of the hermitian vector bundle
$(E,\Dbar_s,H\lsT)$. By our construction, it satisfies
$$\LL_{s,1}(H\lsT,\omega\lsT)\equiv 0\!\!\mod V_1,\quad
\LL_{s,2}(H\lsT,\omega\lsT)=0\and \LL_{s,3}(H\lsT,\omega\lsT)=0.
$$
Hence to find solutions to $\LL_s=0$ it suffices to investigate
the vanishing loci of the functional $\brr(s,\cdot)$ from
$B_\epsilon(T_0)$ to the Lie algebra $\uu(r)$ defined by
\begin{equation}\lab{8.4}
\brr(s,T)=\int_X\bigl[\bL_{s,1}(H\lsT,\omega\lsT)\bigr]_{1},
\end{equation}
where $[\cdot]_1$ is the projection from
$\Omega_\RR^{\bullet}\bl\su E\br$ to $\Omega_\RR^{\bullet}\bl\uu(
\CC_X\ur)\br$. Here $\uu(\CC_X\ur)$ is the bundle of
$<\,,>$-hermitian antisymmetric endomorphisms of $\CC_X\ur$.

As in section two, we shall first prove $\dot \brr(0,T)=0$ for all
$T$. Indeed,
\begin{equation}\lab{4.5}
\dot \brr(0,T)=\int_X T\umhalf\bigl[\dot F_{0,T}\bigr]_1
T\uhalf\wedge\omega_{0,T}^2 +2\int_X T\umhalf\bigl[F_{0,T}\bigr]_1
T\uhalf\wedge\omega_{0,T}\wedge\dot\omega_{0,T}.
\end{equation}
Because $H_{0,T}$ is a direct sum of a flat metric on $\CC_X\ur$
and a metric on $TX$, under the decomposition $E=\CC_X\ut\oplus
TX$,
\begin{equation}\lab{7.3}
F_{0,T}=\left(%
\begin{array}{cc}
  0 & 0 \\
  0 &\ast \\
\end{array}%
\right)\in \Omega^{1,1}_\RR(\su E).
\end{equation}
What we will actually show is that
$$\dot F_{0,T}\wedge\omega\lzT^2=\left(%
\begin{array}{cc}
  0 & 0 \\
  0 &\ast \\
\end{array}%
\right)\in \Omega^{6}_\RR(\su E).
$$
Since $(H\lsT,\omega\lsT)$ are solutions to $\bL_s=0 \mod V_1$,
there is a function $\bc(s,T)$ taking values in $V_1$ with
$\bc(0,T)=0$ so that
$$ F\lsT\wedge \omega\lsT^2=H\lsT\uhalf \bc(s,T) H\lsT\umhalf.
$$
Taking derivative of $s$ at $s=0$, and coupled with $\bc(0,T)=0$,
we obtain
\begin{equation}\lab{7.2}
\dot F\lzT\wedge\omega\lzT^2+2
F\lzT\wedge\omega\lzT\wedge\dot\omega\lzT=H\uhalf\lzT\dot\bc(0,T)
H\umhalf\lzT,
\end{equation}
which, after projecting to $\Omega^6_\RR(\uu(\CC_X\ur))$, becomes
\begin{equation}
\bigl[\dot F\lzT\wedge\omega\lzT^2\bigr]_1=T\uhalf\dot\bc(0,T)
T\umhalf.
\end{equation}

Next, we let $F_s$ as in (\ref{2.11}) be the curvature of
$(E,\Dbar_s,I)$. Because
$$F\lsT=F_s+\Dbar_s(D_s\pri H\lsT\cdot H\lsT\upmo),
$$
because $D_0\pri H\lzT=0$, and because $D_s$ is a direct sum of a
flat connection on $\CC_X\ur$ and a Hermitian Yang-Mills
connection on $TX$,
$$[\dot F\lzT]_1=[\dot F_0]_1+
\Dbar_0[\dot{D}\pri_0 H\lzT\cdot H\lzT\upmo]_1 +\Dbar_0[D\pri_0
\dot H\lzT\cdot H\lzT\upmo]_1.
$$
Using the expression of $F_s$ in (\ref{2.11}), and that $C_{11}=0$
as stated in (\ref{4.9}),
\begin{equation}\lab{7.7}
[\dot F_0]_1=D_0\pri C_{11}-\Dbar_0 C_{11}\sta=0.
\end{equation}
Hence
$$\bigl[ \dot F\lzT\bigr]_1=D_0\pri \varphi_1+ D_0\dpri \varphi_2
$$
for some sections $\varphi_1$ and $\varphi_2$. Therefore, by Lemma
\ref{2.2}
$$\int_X T\uhalf\dot\bc(0,T)
T\umhalf=\int_X \bigl[\dot F\lzT\bigr]_1\wedge\omega\lzT^2=\int_X
\bl D_0\pri \varphi_1+ D_0\dpri \varphi_2\br\wedge\omega\lzT^2=0.
$$
Since $\dot\bc(0,T)/\omega_0^3$ is a constant section of
$\End(\CC_X\ur)$, the above vanishing forces $\dot\bc(0,T)=0$,
which simplifies (\ref{7.2}) to
$$\dot F\lzT\wedge\omega\lzT^2+2
F\lzT\wedge\omega\lzT\wedge\dot\omega\lzT=0.
$$
Finally, because $F\lzT$ has vanishing entries as shown in
(\ref{7.3}),
\begin{equation}\lab{7.4}
\dot F\lzT\wedge\omega\lzT^2= \left(\begin{array}{cc}
  0 & 0 \\
  0 &\ast \\
\end{array}%
\right)\in \Omega^{6}_\RR(\su E).
\end{equation}
The vanishing (\ref{4.5}) follows from (\ref{7.3}) and
(\ref{7.4}).

We next compute $\ddot \brr(0,T)$. First, because $F\lzT=0$,
$$\bl H\lsT\umhalf F\lsT H\lsT\uhalf\cdot
\frac{d^2}{ds^2}\omega\lsT^2\br|_{s=0}=0
$$
Because $\bigl[ \dot F\lzT\bigr]_1=0$,
$$\bigl[ \dds \bl H\lsT\umhalf F\lsT
H\lsT\uhalf\br\wedge\dds \bl\omega\lsT^2\br|_{s=0}\bigr]_1=0.
$$
Hence
$$\bigl[\frac{d^2}{ds^2}\bl H\lsT\umhalf F\lsT H\lsT
\uhalf\wedge\omega\lsT^2\br|_{s=0}\bigr]_1=
\bigl[\frac{d^2}{ds^2}\bl H\lsT\umhalf F\lsT
H\lsT\uhalf\br|_{s=0}\wedge\omega\lzT^2\bigr]_1.
$$
Taking second order derivative of $H\lsT\umhalf F\lsT H\lsT
\uhalf$, we will encounter terms like
$$\frac{d^2}{ds^2}\bl H\lsT\umhalf\br|_{s=0}\cdot F\lzT\cdot H\lzT\uhalf,
$$
which are all zero because $F\lzT=0$. We will also encounter terms
like
$$\dds\bl H\lsT\umhalf\br|_{s=0}\cdot \dot F\lzT\cdot H\lzT\uhalf;
$$
after wedging it with $\omega\lzT^2$, because $H\lzT\uhalf$ is
diagonal and $\dot F\lzT\wedge\omega\lzT^2$ has vanishing shown in
(\ref{7.4}), their projections to $\Omega_\RR^6(\uu(\CC_X\ur))$
are zero also. Hence the only term left is
$$\bigl[\frac{d^2}{ds^2}\bl H\lsT\umhalf F\lsT H\lsT
\uhalf\wedge\omega\lsT^2\br|_{s=0}\bigr]_1= H\lsT\umhalf \ddot
F\lzT H\lzT \uhalf\wedge\omega\lzT^2.
$$

As in the previous section, we compute
$$\int \bigl[H\lzT\umhalf\ddot
F\lzT H\lzT\uhalf\bigr]_1\wedge\omega\lzT^2=
\qquad\qquad\qquad\qquad\qquad\qquad\qquad\qquad\qquad\qquad\qquad
$$
$$\qquad\qquad\quad=
\int_X T\umhalf[\ddot F_0]_1 T\uhalf\wedge\omega\lzT^2 -2\int_X
T\umhalf\bigl[[\dot A_0,[\dot A_0\sta,
\hh\lzT]\hh\lzT\upmo]\bigr]_1 T\uhalf\wedge\omega\lzT^2-\qquad
$$%\end{equation}
$$\qquad\qquad\qquad
-2\int_X T\umhalf\bigl[ [\dot A_0,D\pri_0\dot \hh\lzT\cdot
\hh\lzT\upmo]\bigr]_1T\uhalf\wedge\omega\lzT^2 +\int_X
T\umhalf\bigl[ \Dbar_0\Phi_{T}\bigr]_1T\uhalf\wedge\omega\lzT^2
$$
for some form $\Phi\lzT$. We now look at the four terms in the
above identity: the last term vanishes because of Lemma \ref{2.2};
the next-to-last term is
$$-2\int_X T\umhalf\bigl[\dot A_0\cdot D_0\pri\dot H\lzT\cdot
H\lzT\upmo\bigr]_1T\uhalf\wedge\omega\lzT^2+2\int_X T\umhalf\bigl[
D_0\pri\dot H\lzT\cdot H\lzT\upmo\cdot \dot
A_0\bigr]_1T\uhalf\wedge\omega\lzT^2,
$$
which is zero because $D_0^{\prime\prime\ast} \dot A_0=0$ and
Lemma \ref{2.2}. Using
$$\dot A_0=\left(%
\begin{array}{cc}
  0 & C_{12} \\
  C_{21} & C_{22} \\
\end{array}%
\right) \and
H\lzT =\left(%
\begin{array}{cc}
  T & 0 \\
  0 & \alpha I_2 \\
\end{array}%
\right),\quad \alpha=|T|^{-1/3}
$$
one computes
$$\bigl[ [\dot A_0,D\pri_0\dot \hh\lzT\cdot
\hh\lzT\upmo]\bigr]_1=C_{12}\wedge C_{12}\sta\bl I_1-\alpha
T\upmo\br+(I_1-\alpha\upmo T\br C_{21}\sta\wedge C_{21}.
$$
For the same reason,
$$\bigl[\ddot F_0\bigr]_1\wedge\omega\lzT^2=\bigl[2\dot A_0\wedge\dot A_0\sta
+ 2\dot A_0\sta\wedge\dot A_0\bigr]_1\wedge\omega\lzT^2 =2\bl
C_{12}\wedge C_{12}\sta+C_{21}\sta\wedge
C_{21}\br\wedge\omega\lzT^2.
$$
Therefore,
$$\ddot\brr(0,T)=2\int_X\bl\alpha T\umhalf C_{12}\wedge C_{12}\sta
T\umhalf +\alpha\upmo T\uhalf C_{21}\sta\wedge C_{21} T\uhalf\br
\wedge\omega\lzT^2.
$$

We now investigate the solvability of $\brr(s,T)=0$ for small $s$.
For this, we need to make an assumption on the class $C_{12}$ and
$C_{21}$. Recall that $C_{12}$ is a column vector
$[\alpha_1,\cdots,\alpha_r]^t$ whose components are
$\Dbar_0$-harmonic
$$\alpha_i\in\Omega^{0,1}\bl TX\dual\otimes\CC_X\br;
$$
$C_{21}$ is a row vector $[\beta_1,\cdots,\beta_r]$ whose
components are $\Dbar_0$-harmonic
$$\beta_i\in\Omega^{0,1}\bl\CC_X\dual\otimes TX\br.
$$
Since both $\alpha_i$ and $\beta_i$ are (1,0)-forms,
\begin{equation}
\sqrto B=\sqrto\int_X C_{12}\wedge
C_{12}\sta\wedge\omega\lzT^2\and \sqrto B\pri=-\sqrto\int_X
C_{21}\sta\wedge C_{21}\wedge\omega\lzT^2
\end{equation}
are non-negative definite hermitian symmetric matrices. Because
$\alpha_i$ are $\Dbar_0$-harmonic, $\sqrto B$ is positive definite
if and only if $[\alpha_1],\cdots,[\alpha_r]$ are linearly
independent elements in $H^1_{\dbar}(TX\dual)$. Similarly, $\sqrto
B\pri$ is positive definite if $[\beta_1],\cdots,[\beta_r]$ are
linearly independent in $H^1_{\dbar}(X, TX)$. Hence the positivity
of $\sqrto B$ and $\sqrto B\pri$ only depend on the
Kodaira-Spencer class $\kappa\in H^1_{\dbar}(X,E\dual\otimes E)$.

We now assume that {\sl both matrices $\sqrto B$ and $\sqrto
B\pri$ are positive definite.} By a GL(r,$\CC$) change of basis of
$\CC_X\ur$, we can assume that $\sqrto B\pri=I_{r\times r}$. Then
$$\ddot\brr(0,T)=2|T|^{-1/3}T\umhalf B T\umhalf+2\sqrto|T|^{1/3}T
\in \uu(r).
$$
Clearly, $\ddot\brr(0,T)=0$ if $T$ is
$$T_0\triangleq|\sqrto B_1|^{{1}/{2(r+3)}}\bl\sqrto B\br^{{1}/{2}}.
$$

\begin{lemm}
The map $\Phi\mh \partial\epball\to S(1)$ to the unit sphere
$S(1)\sub \uu(r)$ defined by
$$\Phi(T)= \frac{\ddot\brr(0,T)}{\|\ddot\brr(0,T)\|}$$
is a degree one map.
\end{lemm}

\begin{proof}
We define
$$\bu_t(T)=|T|^{-1/3}\bl tT+(1-t)I\br\umhalf \bl B
+\sqrto|T|^{1/3}T^2 \br\bl tT+(1-t)I\br\umhalf
$$
and consider
$$\frac{\bu_t(\cdot)}{\|\bu_t(\cdot)\|}: \partial\epball\to S(1).
$$
It is well-defined since $T$ and $I$ are positive definite; it is
$\Phi$ when $t=1$. Hence it provides a homotopy between $\Phi$ and
$$\Phi_1(\cdot)=\frac{\bu_0(\cdot)}{\|\bu_0(\cdot)\|}: \partial\epball\to S(1).
$$
Next we consider
$$\bv_t(T)=\bB+\sqrto\bl(1-t)|T|^{2/3}+t|T_0|^{2/3}\br T^2.
$$
We claim that $\bv_t(T)\ne 0$ for all $t\in [0,1]$. Suppose for
some $t_0\in [0,1]$ and $T\in\partial\epball$,
$$\bB+\sqrto\bl(1-t_0)|T|^{2/3}+t_0|T_0|^{2/3}\br T^2=0,
$$
then $T=\eta (\sqrto B)^{1/2}$ for some $\eta\in \RR^+$. Since
$T\in \partial B_\epsilon(T_0)$, $\eta$ satisfies
$$\|T-T_0\|=|\eta-|\sqrto B|^{-1/2(r+3)}|\,\|(\sqrto B)^{1/2}\|=\epsilon.
$$
Hence $\eta$ can only take values
$$\eta_\pm=|\sqrto B|^{-1/2(r+3)}\pm\epsilon\pri,\quad
\epsilon\pri=\epsilon/\|(\sqrto B)^{1/2}\|.
$$
But then $|\eta_+(\sqrto B)^{1/2}|>|\sqrto B|^{3/2(r+3)}=|T_0|$;
and then
$$\bl t_0|\eta_+(\sqrto B)^{1/2}|^{2/3}+(1-t_0)|T_0|^{2/3}\br \eta_+^2>
\qquad\qquad\qquad\qquad\qquad
$$
$$\qquad\qquad\qquad\qquad\qquad
>\bl t_0|T_0|^{2/3}+(1-t_0)|T_0|^{2/3}\br\bl|\sqrto
B|^{-1/2(r+3)}+\epsilon\pri\br^2>1.
$$
Hence $\bv_{t_0}(\eta_+(\sqrto B)^{1/2})\ne 0$. Similarly,
$\bv_{t_0}(\eta_-(\sqrto B)^{1/2})\ne 0$. This proves that
$$\frac{\bv_t(\cdot)}{\|\bv_t(\cdot)\|}:\partial\epball\lra S(1)
$$
are well-defined and is a homotopy between $\Phi_1$ and
$$\Phi_2: \partial\epball\lra S(1);\quad \Phi_2(T)=
\frac{\bB+\sqrto|T_0|^{2/3} T^2}{\| \bB+\sqrto|T_0|^{2/3} T^2\|}.
$$

It remains to show that $\deg\Phi_2=1$. We write $T=T_0+\epsilon
\Delta T$ with $\Delta T$ varies in the unit sphere in the space
of hermitian symmetric matrices $\cH_{r\times r}$. Under this form
the numerator of $\Phi_2$ is
$$\bB+\sqrto|T_0|^{2/3} (T_0+\epsilon\Delta T)^2=\sqrto|T_0|^{2/3}\bl
\Delta T T_0+T_0\Delta T\br+\epsilon^2 |T_0|^{2/3}(\Delta T)^2.
$$
For $\epsilon$ small enough, the degree of $\Phi_2$ is the same as
the degree of
\begin{equation}\lab{6.1}
\Delta T\longmapsto \sqrto\frac{\Delta TT_0+T_0\Delta T}{\|\Delta
TT_0+T_0\Delta T\|},
\end{equation}
which is the same as
$$\Delta T\longmapsto \sqrto\Delta T.
$$
Because the map $\partial B_1(0)\sub\cH_{r\times r}\to S(1)\sub
\uu(r)$ by multiplying $\sqrto$ has degree one, the map $\Phi$ has
degree one as well. This proves the Lemma.
\end{proof}

We are now ready to prove the theorem

\begin{theo}\lab{p4.2}
Let $(X,\omega_0)$ be a Calabi-Yau threefold; let $\Dbar_s$ be a
smooth deformation of the tautological holomorphic structure
$\Dbar_0$ on $E=\CC_X\ur\oplus TX$. Suppose the Kodaira-Spencer
class $\kappa\in H^1_{\dbar}(X,E\dual\otimes E)$ of the family
$\Dbar_s$ at $s=0$ satisfies the non-degeneracy condition that
both $\sqrto B$ and $\sqrto B\pri$ in (\ref{6.1}) are positive
definite. Then for sufficiently large $c\in \RR$ and small $a>0$,
there is a family of pairs of hermitian metrics and hermitian
forms $(H_s,\omega_s)$, not necessarily continuous in $s\in
[0,a)$, so that
\newline
1. the $\omega_0$-harmonic part of $\omega_s$ is $c\omega_0$;
\newline
2. the pair $(H_s,\omega_s)$ is a solution to Strominger's system
for the holomorphic vector bundle $(E,\Dbar_s)$;
\newline
3. $\lim_{s\to 0} \omega_s=c\omega_0$; $\lim_{s\to 0} H_s$ is a
Hermitian Yang-Mills connection of $E$ over $(X,\omega_0)$.
\end{theo}

\begin{proof}
First, we pick a basis of $\CC_X\ur$ so that the matrix $\sqrto
B\pri$ in (\ref{6.1}) is the identity matrix. We let $B$ be the
other matrix and let $T_0=|\sqrto B|^{1/2(r+3)}(\sqrto B)^{1/2}$.
By Lemma \ref{6.3}, we can choose $C$ so that Lemma \ref{6.3}
holds for $T_0$ chosen. Then for any $c>C$, we form solution set
$\cS_s$ of the system $\bP\circ\bL_s=0$ and parameterize the
solutions near $(H_{T_0},c\omega_0)$ by $(s,T)\in
[0,a)\times\epball$. Based in this parameterization, we then form
the functional $\brr(s,T)$ in (\ref{8.4}). Because
$\dot\brr(0,T)=0$ and
\begin{equation}\lab{8.1}
\frac{\ddot\brr(0,\cdot)}{\|\ddot\brr(0,\cdot)\|}:
\partial\epball\lra S(1)
\end{equation}
has degree one, for some small $0<a\pri<a$ the maps
$$\brr(s,\cdot): \partial\epball\lra \uu(r), \quad s\in(0,a\pri)
$$
does not take the value $0\in\uu(r)$. Hence the associated map
\begin{equation}\lab{8.2}
\frac{\brr(s,\cdot)}{\|\brr(s,\cdot)\|}:\partial\epball\lra
S(1)\sub\uu(r),\quad s\in(0,a\pri),
\end{equation}
has the same degree as that of (\ref{8.1}), which is one. Hence
the map
$$\brr(s,\cdot): \epball\lra \uu(r), \quad s\in(0,a\pri),
$$
attains value $0\in\uu(r)$ for all $s\in (0,a\pri)$ in
$B_\epsilon(T_0)$. This proves the first two part of the theorem.
The last part is true because we can choose $\epsilon$ arbitrarily
small.
\end{proof}

\section{Irreducible Solutions on quintic threefolds}

So far we have derived a sufficient condition for the existence of
irreducible solutions to Strominger's system. Our next step is to
find examples that satisfy this condition. It is the purpose of
this section to work out examples for SU(4) and SU(5).

We will first consider the Fermat quintic
$$X=\{z_0^5+z_1^5+z_2^5+z_3^5+z_4^5=0\}\sub\Pfour;
$$
we will find a deformation of the holomorphic structure of
$\CC_X\oplus TX$ and show that it satisfies the requirement of
theorem \ref{p4.2}. This will provide us SU(4) solutions to
Strominger's system.

We begin with the Euler exact sequence of $T\Pfour$ (the middle
column), and the exact sequence relating $TX$ and the restriction
to $X$ of the tangent bundle $T_X\Pfour=T\Pfour|_X$ (the top row):
\begin{equation}\lab{9.8}
\begin{CD}
@.0@. 0\\
@. @AAA @AAA\\
0 @>>>TX @>{\varphi_1}>> T_X\Pfour @>{\varphi_2}>>\cO_X(5) @>>> 0\\
@. @AAA @AAA @|\\
0 @>>>F @>>> \cO_X(1)^{\oplus 5} @>>>\cO_X(5) @>>> 0\\
@. @AAA @AAA\\
@.\cO_X @= \cO_X\\
@. @AAA @AAA\\
@.0@. 0\\
\end{CD}
\end{equation}
We take $F$ be the kernel of $\cO_X(1)^{\oplus 5}\lra\cO_X(5)$ and
fill in the remainder entries to make up the exact diagram as
shown above.

We claim that the left column in (\ref{9.8}) is non-split. Assume
not, say $F=TX\oplus\cO_X$. Then since $F$ is a subsheaf of
$\cO_X(1)\uf$ with quotient sheaf $\cO_X(5)$, $\cO_X(1)\uf/TX$
must be locally free and an extension of $\cO_X(5)$ by $\cO_X$.
Because $\Ext_X^1(\cO_X(5),\cO_X)=0$, the only extension of
$\cO_X(5)$ by $\cO_X$ is the direct sum $\cof\oplus\cO_X$. Hence
$$\coo\uf/TX\cong \cO_X\oplus\cO_X(5).
$$
In particular, $\cO_X$ becomes a quotient sheaf of $\coo\uf$ that
is impossible. This proves that it does not split.

Next, we will construct a deformation of holomorphic structure of
$\CC_X\oplus TX$ so that its Kodaira-Spencer class is of the form
\begin{equation}\lab{9.1}
\kappa=\left(%
\begin{array}{cc}
  0 & 0 \\
  \xi & 0 \\
\end{array}%
\right) \in\Ext_X^1(\cO_X\oplus TX,\cO_X\oplus TX)
\end{equation}
whose only non-trivial entry is the extension class
$\xi\in\Ext^1_X(TX,\cO_X)$ of the left column exact sequence in
(\ref{9.8}); $\xi$ is non-trivial because the exact sequence does
not split. We let
$$\pi_1:X\times\Ao\lra X\and\pi_2:X\times\Ao\lra \Ao
$$
be the projections; we let $t$ be the standard coordinate function
on $\Ao$. The class
$$t\cdot\xi\in
\Gamma(\cO_\Ao)\otimes\Ext^1_X(TX,\cO_X)=\Ext^1_{X\times\Ao}\bl
\pi_1\sta TX,\cO_{X\times\Ao}\br
$$
defines an extension sheaf over $X\times\Ao$:
\begin{equation}\lab{5.2}
0\lra \cO_{X\times\Ao}\lra \cF\lra \pi_1\sta TX\lra 0.
\end{equation}
The extension sheaf $\cF$ is locally free; its restriction to
$X\times t$, which we denote by $F_t$, form a one parameter family
of holomorphic vector bundles whose special member $F_0\cong
\cO_X\oplus TX$ and its general member $F_t\cong F$ for $t\ne 0$.
Here by abuse of notation we use $t$ to denote the point in $\Ao$
having coordinate $t$. It is a tautology that the Kodaira-Spencer
class of this family at $t=0$ is the $\kappa$ in (\ref{9.1}).

In terms of differential geometry, if we fix smooth isomorphisms
$F_t\cong \CC_X\oplus TX$ that also depend smoothly on $t$, then
the holomorphic structure on $F_t$ induces a family of holomorphic
structures $\Dbar_t$ on $E=\CC_X\oplus TX$ that is a deformation
of the holomorphic structure $\Dbar_0$ on $\CC_X\oplus TX$.
Following the convention of the first part of this paper, if we
write $\Dbar_t=\Dbar_0+A_t$ and use the splitting $E=\CC_X\oplus
TX$, then
$$\dot D_0^{\prime\prime}=\dot A_0=\left(%
\begin{array}{cc}
  0 & 0 \\
  C_{21} & 0 \\
\end{array}%
\right)
$$
and $C_{21}$ represents the class $\xi$ in $H^1(TX\dual)$; thus
$[C_{21}]\ne 0$.

What we aim at is to find a deformation of holomorphic structures
$\Dbar_t$ of $(E,D\dpri_0)$ so that the first order deformation
$$\dot D_0^{\prime\prime}=\left(%
\begin{array}{cc}
  0 & C_{12} \\
  C_{21} & C_{22} \\
\end{array}%
\right)
$$
will have $[C_{12}]\ne0$ and $[C_{21}]\ne0$. To achieve this, we
will construct a smooth family of holomorphic vector bundles
$(E,\Dbar_u)$ parameterized by a smooth pointed domain $0\in U$ so
that
\begin{enumerate}
\item $\Dbar_0$ is the holomorphic structure on $\CC_X\oplus TX$;

\item there is a path $u=\rho_1(t)$ in $U$ with $\rho_1(0)=0$ so
that
$%\displaystyle
\dot D_{\rho_1(0)}^{\prime\prime}=\left(%
\begin{array}{cc}
  0 & \ast \\
  C_{21} & \ast \\
\end{array}%
\right) $ and $[C_{21}]\ne 0$;

\item there is another path $u=\rho_2(t)$ in $U$ with
$\rho_2(0)=0$ so that
$%\displaystyle
\dot D_{\rho_2(0)}^{\prime\prime}=\left(%
\begin{array}{cc}
  0 & C_{12} \\
  \ast & \ast \\
\end{array}%
\right)$ and $[C_{12}]\ne 0$.
\end{enumerate}

As we saw before, for the first path all we need is to have it
represent the family $F_t$ constructed in (\ref{5.3}). We now
construct the second family that will represent the path $\rho_2$
that we need. We will work out the family over $U$ after we have
done this.

Using the top row exact sequence of the diagram (\ref{9.8}), we
can fit $\cO_X\oplus TX$ into the exact sequence
\begin{equation}\lab{5.5}
0\lra \cO_X\oplus TX\lra \cO_X\oplus T_X\Pfour\mapright{\varphi}
\cof\lra 0.
\end{equation}
Here $\varphi=(0,\varphi_2)^t$ is $0$ when restricted to $\cO_X$,
and is the $\varphi_2$ in the diagram when restricted to
$T_X\bP^4$. We then pick a section $u\in H^0(\cof)$, viewed as a
homomorphism $\cO_X\to\cof$, to form a new homomorphism of sheaves
over $X\times\Ao$:
$$\Phi=(tu,\pi_1\sta\varphi_2)^t:
\cO_{X\times\Ao}\oplus \pi_1\sta T_X \Pfour \mapright{\Phi}
\pi_1\sta\cof
$$
whose restriction to $\cO_{X\times\Ao}$ (resp. $\pi_1\sta
T_X\Pfour$) is $tu$ (resp. $\pi_1\sta\varphi_2$). We let $\cF\pri$
be the kernel of $\Phi$. $\cF\pri$ fits into the middle rwo exact
sequence
\begin{equation}\lab{5.6}
\begin{CD}
@.0@. 0\\
@.@VVV@VVV\\
@.\pi_1\sta TX@= \pi_1\sta TX\\
@.@VV{\Psi}V@VV{(0,\pi_1\sta\varphi_1)}V\\
0@>>> \cF\pri @>>> \cO_{X\times\Ao}\oplus \pi_1\sta T_X \Pfour
@>{\Phi}>> \pi_1\sta\cof @>>> 0\\
@. @VVV @VVV @|\\
0 @>>> \cO_{X\times\Ao} @>>>
\cO_{X\times\Ao}\oplus\pi_1\sta\cO_X(5) @>>> \pi_1\sta\cO_X(5)
@>>> 0\\
@. @VVV @VVV \\
@. 0@.0\\
\end{CD}
\end{equation}
Because the composite
$$\Phi\circ(0,\pi_1\sta\varphi_1)=0,
$$
$(0,\pi_1\sta\varphi_1)$ lifts to $\Psi$, shown in the diagram;
its cokernel is $\cO_{X\times\Ao}$.

We denote the restriction to $X\times\{t\}$ of $\cF\pri$ by
$F\pri_t$. Clearly, $F_0\pri\cong\cO_X\oplus TX$. The
Kodaira-Spencer class of the first order deformation of the family
$\cF\pri$ at $t=0$ is
$$\kappa\pri=\left(%
\begin{array}{cc}
  0 & \kappa_{12}\pri \\
  0 & 0 \\
\end{array}%
\right).
$$
To show that $\cF\pri$ is the desired family we need to show that
$\kappa_{12}\pri\ne 0$. We now prove that this is true. We let
$\bA_2=\spec\CC[t]/(t^2)$, which in plain language is the first
order infinitesimal neighborhood of $0\in\Ao$. Suppose
$\kappa_{12}\pri=0$, then based on deformation theory of vector
bundles, the induced sheaf homomorphism
$$\psi_2: \cF\pri\otimes_{\cO_{X\times\Ao}}\cO_{X\times\bA_2}\lra
\cO_{X\times\bA_2},\ \text{or equivalently}\
\cF\pri|_{X\times\bA_2}\to \CC_{X\times\bA_2},
$$
splits. Namely, there is a homomorphism
\begin{equation}\lab{5.8}
\tilde\psi_2: \cO_{X\times\bA_2}\lra
\cF\pri\otimes_{\cO_{X\times\Ao}}\cO_{X\times\bA_2}
\end{equation}
so that
$$\psi_2\circ\tilde\psi_2=\text{id}.
$$
Let $p\mh X\times\bA_2\to X$ be the projection. Since $\cF\pri$ is
defined by the exact sequence (\ref{5.6}), the homomorphism
$\tilde\psi_2$ induces a homomorphism
$$\cO_{X\times\bA_2}\lra \cO_{X\times\bA_2}\oplus
p\sta T_X\bP^4;
$$
because $\Ext^1_X(\cO_X,\cO_X)=0$ it lifts to a
$$\mu: \cO_{X\times\bA_2}\lra \cO_{X\times\bA_2}\oplus
p\sta \cO_{X}(1)\uf.
$$
Let
$$\lambda:\cO_{X\times\bA_2}\oplus
p\sta\cO_{X}(1)\uf\lra p\sta\cO_{X}(5)
$$
be the restriction of the composite of
$$\cO_{X\times\Ao}\oplus
p\sta \cO_{X}(1)\uf \lra \cO_{X\times\Ao}\oplus p\sta T_X\bP^4
$$
and $\Phi$ in (\ref{5.6}) to $X\times\bA_2$. Then by definition
$$\lambda\circ\mu=0.
$$
To study this identity, we notice that the homomorphism $\mu$ must
be of the form
$$\mu=[1+at,bz_0+t\alpha_0,\cdots,bz_4+t\alpha_4]
$$
with $[z_0,\cdots,z_4]$ the homogeneous coordinate of $\bP^4$,
$\alpha_i\in H^0(\cO_X(1))$, $a\in \CC$ and $b\in \CC[t]/(t^2)$;
the homomorphism $\lambda$ is of the form
$$\left[%
\begin{array}{c}
  tu \\ z_0^4\\ \vdots \\ z_4^4
\end{array}%
\right]
$$
Because $\lambda\circ\mu=0$ holds over $X\times \bA_2$, we have
$$[1+a_1t,bz_0+t\alpha_0,\cdots,bz_4+t\alpha_4]\left[%
\begin{array}{c}
  tu \\ z_0^4\\ \vdots \\ z_4^4
\end{array}%
\right]\equiv 0\mod(t^2, z_0^5+\cdots+z_4^5).
$$
After simplification, the above identity reduces to
$$u+\alpha_0 z_0^4+\cdots+\alpha_4 z_4^4\equiv
0\mod(z_0^5+\cdots+z_4^5).
$$
Now we choose $u=z_0^2z_1^3$. It is clear that there are no
$\alpha_i\in H^0(\cO_X(1))$ that make the above identity holds.
Hence with such choice of $u$ the lift $\tilde\psi_2$ does not
exist. This proves $\kappa_{12}\pri\ne 0$.

It remains to find a family of holomorphic vector bundles that
includes the two families $\cF$ and $\cF\pri$ as its subfamilies.
We let $\eta\in \Ext_X^1(T_X\bP^4,\cO_X)$ be the extension class
of the Euler exact sequence
\begin{equation}\lab{5.3}
0\lra\cO_X\lra\cO_X(1)\uf\lra T_X\bP^4\lra 0.
\end{equation}
Then $t\eta$ is an extension class
$$t\eta\in \Gamma(\cO_\Ao)\otimes\Ext^1(T_X\bP^4,\cO_X)=
\Ext_{X\times\Ao}^1(\pi_1\sta T_X\bP^4,\cO_{X\times\Ao})
$$
that defines an exact sequence over $X\times\Ao$:
$$0\lra\cO_{X\times\Ao}\lra\cW\lra \pi_1\sta T_X\bP^4\lra 0.
$$
Clearly, $\cW\otimes_{\cO_{X\times\Ao}}\cO_{X\times
\{0\}}=\cO_X\oplus T_X\bP^4$ while
$\cW\otimes_{\cO_{X\times\Ao}}\cO_{X\times \{t\}}=\coo\uf$ for
$t\ne 0$. We claim that
\begin{equation}\lab{5.11}
\pi_{2\ast}\bl \cW\dual\otimes\pi_1\sta\cof\br
\end{equation}
is a locally free sheaf of $\cO_{\Ao}$-modules. By base change
property, this is true if
$$H^1(X,\bl \cO_X\oplus T_X\bP^4)\dual\otimes\cof\br=0
\and H^1\bl X,(\coo\uf)\dual\otimes\cof\br=0.
$$
Since $X\sub\bP^4$ is a smooth hypersurface, a standard long exact
sequence chasing shows that $H^1(X,\cO_X(a))=0$ for any integer
$a$. To prove the above two identities, we only need to check that
$H^1(X,T_X\dual\bP^4\otimes\cof)=0$. For this, we apply the long
exact sequence of cohomologies to the dual of (\ref{5.3}) tensored
with $\cof$:
$$H^0(\cO_X(4)\uf)\lra H^0(\cof)\lra H^1(T_X\dual\bP^4(5))\lra
H^1(\cO_X(4)\uf).
$$
Because the last term is zero, and because $H^0(\cO_{\bP^4}(a))\to
H^0(\cO_X(a))$ is surjective, the first arrow is surjective. This
shows that $H^1(T_X\dual\bP^4(5))=0$, and hence (\ref{5.11}) is
locally free.

We now let $W$ be the total space of the vector bundle
(\ref{5.11}) and let
$$q\mh X\times W\to X\times \Ao
$$
be the projection. Over $X\times W$ there is a tautological
homomorphism
$$q\sta \cW\lra q\sta\pi_1\sta\cof.
$$
Let $\cE$ be the kernel of the above sheaf homomorphism; for $w\in
W$ we denote by $E_w$ the restriction of $\cE$ to $X\times w$.

It is now a matter of direct checking that there are two paths
$\rho_1(t)$ and $\rho_2(t)$ in $W$ so that $E_{\rho_1(t)}$ and
$E_{\rho_2(t)}$ represent $F_t$ and $F\pri_t$ respectively. First
of all, the homomorphism $\varphi\mh \cO_X\oplus T_X\bP^4\to\cof$
in (\ref{5.5}) represents a point in $W$; we designate this point
to be the marked point $0\in W$. The family $\cF\pri$ is
constructed as the kernel of $\Phi$ in (\ref{5.6}) with $\Phi$
restricting to $X\times\{0\}$ being $\varphi$. Hence $\Phi$
represents a path $\rho_2$ in $W$ initiating from $0$ and is
contained in the fiber of $W\to\Ao$ over $0\in\Ao$ that satisfies
$E_{\rho_2(t)}\cong F_t\pri$.

As to the first family $\cF$ constructed in (\ref{5.3}), it fits
into the exact diagram
$$\begin{CD}
@.@.0@.0\\
@.@.@VVV@VVV\\
0@>>> \cO_{X\times\Ao} @>>> \cF @>>> \pi_1\sta TX @>>> 0\\
@.@|@VVV@VVV\\
0 @>>> \cO_{X\times\Ao} @>>> \cW @>>> \pi_1\sta T_X\Pfour @>>> 0\\
@.@.@VV{\Psi}V @VVV\\
@.@.\pi_1\sta\cO_X(5)@=\pi_1\sta\cO_X(5)\\
@.@.@VVV@VVV\\
@.@.0@.0\\
\end{CD}
$$
Since $\Psi$ restricting to $X\times\{0\}$ is the $\varphi$ in
(\ref{5.5}), it represents a path $\rho_1$ in $W$ with
$\rho_1(0)=0$ so that $E_{\rho_1(t)}$ is the first family $F_t$
constructed before.

From what we know of the families $F_t$ and $F_t\pri$, their
Kodaira-Spencer classes at $t=0$ are of the form
$$\left(%
\begin{array}{cc}
  0 & 0 \\
  \kappa_{21} & 0 \\
\end{array}%
\right) \and
\left(%
\begin{array}{cc}
  0 & \kappa_{12}\pri \\
  0 & 0 \\
\end{array}%
\right),\qquad \kappa_{21}\ne 0,\quad \kappa_{12}\pri\ne 0.
$$
Since $W$ is smooth, there is a path $\rho(t)$ with $\rho(0)=0$ so
that $\dot\rho(0)=\dot\rho_1(0)+\dot\rho_2(0)$; hence the family
$E_{\rho(t)}$ has Kodaira-Spencer class at $t=0$
$$\left(%
\begin{array}{cc}
  0 & \kappa_{12}\pri \\
  \kappa_{21} & 0 \\
\end{array}%
\right),\qquad \kappa_{21}\ne 0,\quad \kappa_{12}\pri\ne 0.
$$
It satisfies the requirement of theorem \ref{p4.2}. This proves

\begin{theo}\lab{6.10}
Let $X\sub\bP^4$ be a smooth quintic threefold and $\omega$ is a
Calaby-Yau form (metric) on $X$. Then there is a smooth
deformation $\Dbar_s$ of $(E,\Dbar_0)=\CC_X\oplus TX$ so that for
large $c>0$ and small $s$ there are irreducible regular solutions
$(H_s,\omega_s)$ to Strominger's system on the vector bundle
$(E,\Dbar_s)$ so that $\lim_{s\to 0}\omega_s=c\omega$ and
$\lim_{s\to 0}H_s$ is a regular Hermitian Yang-Mills connection on
$\CC_X\oplus TX$.
\end{theo}

%ch5.tex

We next state the existence of solutions to SU(5)-strominger's
system.

\begin{theo}
Let $X\sub\bP^3\times\bP^3$ be a smooth Calabi-Yau threefold cut
out by three homogeneous polynomials of bi-degrees $(3,0)$,
$(0,3)$ and $(1,1)$. Let $\omega$ be a Calabi-Yau form on $X$.
Then there is a smooth deformation $\Dbar_s$ of
$(E,\Dbar_0)=\CC_X^{\oplus 2}\oplus TX$ so that for large $c>0$
and small $s$ there are irreducible regular solution
$(H_s,\omega_s)$ to Strominger's system on $(E,\Dbar_s)$.
\end{theo}

\begin{proof}
We only need to produce a deformation of holomorphic structure of
$\CC_X\ut\oplus TX$. Let $\pi_1$ and $\pi_2\mh X\to \bP^3$ be the
composite of the immersion $X\sub\bP^3\times\bP^3$ with the
projections $\bP^3\times\bP^3\to\bP^3$. Then $TX$ fits into the
exact sequence
\begin{equation}\lab{6.9}
0\lra TX\lra\pi_1\sta\T\bP^3\oplus\pi_2\sta T\bP^3\lra
\cO_X(3,0)\oplus\cO_X(0,3)\oplus\cO_X(1,1)\lra 0.
\end{equation}
Here $\cO_X(i,j)$ is the restriction to $X$ of
$\pi_1\sta\cO_{\bP^3}(i)\otimes\pi_2\sta\cO_{\bP^3}(j)$. Composing
the canonical
$$\cO_X(1,0)^{\oplus 4}\oplus \cO_X(0,1)^{\oplus
4}\lra \pi_1\sta\T\bP^3\oplus\pi_2\sta T\bP^3
$$
with the last arrow in (\ref{6.9}), we obtain a surjective
$$\cO_X(1,0)^{\oplus 4}\oplus \cO_X(0,1)^{\oplus
4}\mapright{\varphi_2} \cO_X(3,0)\oplus\cO_X(0,3)\oplus\cO_X(1,1)
$$
whose kernel, denoted by $F_0$, is an extension of $TX$ by
$\cO_X\ut$. Next we varies $\varphi_2$ to produce a variation of
holomorphic structure of $F_0$. The bundle $F_0$ is a small
deformation of $\CC_X^{\oplus 2}\oplus TX$; varying $\varphi_2$
produces small deformation of $F_0$. We then mimic the argument in
the proof of Theorem \ref{6.10} to show that we can make this
small deformation of small deformation into a single small
deformation; it is our desired $\Dbar_s$.

To complete the proof of the theorem, we need to check the
non-degeneracy condition on the two matrices $B$ and $B\pri$
associated to the Kodaira-Spencer class $\kappa$ of this family.
It is routine shall be omitted. This completes the proof of the
Theorem.
\end{proof}

\end{document}